\documentclass[pra,showpacs]{revtex4}
\usepackage{graphicx}
\usepackage{subfigure}
\usepackage{caption2}
\usepackage{float}
\usepackage{amsmath}
\usepackage{amssymb}
\usepackage{color}

\begin{document}

\title{Symmetric and asymmetric solitons in linearly coupled Bose-Einstein
condensates trapped in optical lattices}
\author{ Arthur Gubeskys and Boris A. Malomed}
\affiliation{Department of Interdisciplinary Studies, School of Electrical Engineering\\
Faculty of Engineering, Tel Aviv University\\
Tel Aviv 69978, Israel}

\begin{abstract}
We study spontaneous symmetry breaking in a system of two parallel
quasi-one-dimensional traps (\textit{cores}), equipped with optical lattices
(OLs) and filled with a Bose-Einstein condensate (BEC). The cores are
linearly coupled by tunneling (the model may also be interpreted in terms of
spatial solitons in parallel planar optical waveguides with a periodic
modulation of the refractive index). Analysis of the corresponding system of
linearly coupled Gross-Pitaevskii equations (GPEs) reveals that spectral
bandgaps of the single GPE split into subgaps. Symmetry breaking in
two-component BEC solitons is studied in cases of the attractive (AA) and
repulsive (RR) nonlinearity in both traps; the mixed situation, with
repulsion in one trap and attraction in the other (RA), is considered too.
In all the cases, stable asymmetric solitons are found, bifurcating from
symmetric or antisymmetric ones (and destabilizing them), in the AA and RR
systems, respectively. In either case, bi-stability is predicted, with a
nonbifurcating stable branch, either antisymmetric or symmetric, coexisting
with asymmetric ones. Solitons destabilized by the bifurcation tend to
rearrange themselves into their stable asymmetric counterparts. In addition
to the fundamental solitons, branches of twisted (odd) solitons in the AA
system, and twisted bound states of fundamental solitons in both AA and RR
systems, are found too. The impact of a phase mismatch, $\Delta $, between
the OLs in the two cores is also studied. It is concluded that $\Delta =\pi
/2$ only mildly deforms the picture, while $\Delta =\pi $ changes it
drastically, replacing the symmetry-breaking bifurcations by \textit{%
pseudo-bifurcations}, with the branch of asymmetric solutions asymptotically
approaching its symmetric or antisymmetric counterpart (in the AA and RR
system, respectively), rather than splitting off from it. Also considered is
a related model, for a binary BEC in a single-core trap with the OL,
assuming that the two species (representing different spin states of the
same atom) are coupled by linear interconversion. In that case, the
symmetry-breaking bifurcations in the AA and RR models switch their
character, if the inter-species nonlinear interaction becomes stronger than
the intra-species nonlinearity.
\end{abstract}

\pacs{03.75.Lm, 05.45.Yv, 42.65.Tg, 47.20.Ky}
\maketitle

\section{Introduction and the model}

Optical lattices (OLs), i.e., periodic potentials induced by the
interference of counterpropagating coherent laser beams, is a
powerful tool that allows one to support various dynamical
patterns in Bose-Einstein condensates (BECs) \cite{Markus}. In
particular, OLs support \textit{gap solitons} in BEC with
repulsive interaction between atoms (the general topic of BEC
solitons was reviewed in Ref. \cite{Abd}). Gap solitons in BEC
were predicted theoretically \cite{GS,Ostrovskaya} and then
created experimentally in an effectively one-dimensional
(``cigar-shaped") trap equipped with the OL in the axial
direction. In a self-attractive medium, the periodic OL potential
makes it possible to localize a soliton at a prescribed spot, and
supports multi-soliton complexes, as was first demonstrated in a
model of spatial optical solitons in a planar waveguide with a
periodic transverse modulation of the refractive index
\cite{Wang}. Similar theoretical results were then reported for
solitons in self-attractive BECs \cite{Konotop}. In particular, in
the limit case of a very strong OL, the corresponding
Gross-Pitaevskii equation (GPE) \cite{Pit} reduces to a discrete
nonlinear Schr\"{o}dinger (NLS) equation \cite{Smerzi}.

Another topic of great interest to BEC \cite{double-well} and
nonlinear optics \cite{double-well-optics} is \textit{spontaneous
symmetry breaking} of nonlinear fields trapped in dual-core traps
(alias double-well potentials). In particular, the spontaneous
transition from symmetric solitons in the dual-core trap filled
with self-attractive condensate to asymmetric solitons is, in the
first approximation, tantamount to the formation of asymmetric
solitons in dual-core nonlinear optical fibers \cite{dual-core}.
In terms of nonlinear optics, several other dual-core settings
have been studied theoretically, including ones with quadratic
\cite{chi2} and cubic-quintic \cite{Lior} nonlinearities.

Our objective is to explore symmetric, antisymmetric and
asymmetric families of BEC solitons in the model of a dual-core
trap with cores coupled by linear tunneling and equipped with OLs.
We consider symmetric systems with attractive or repulsive
nonlinearity in each core (to be referred to as AA and RR ones,
i.e., ``attractive-attractive" and ``repulsive-repulsive"). In the
AA system, the solitons originate in the semi-infinite bandgap of
the OL's linear spectrum, while in the RR system there are gap
solitons originating in finite bandgaps (we will consider the
first two gaps, demonstrating that the linear coupling between the
cores splits them into \textit{subgaps}). In some cases, soliton
families extend into Bloch bands separating the gaps, thus becoming \textit{%
embedded solitons} \cite{embedded}.

In optics, spontaneous symmetry breaking in two-component gap solitons was
studied in models of dual-core \cite{Mak} and triple-core \cite{Arik} fiber
Bragg gratings (solitons in triangular and planar triple-core optical
waveguides without gratings were considered in Refs. \cite{Buryak}). A mixed
RA system, with repulsion in one core and attraction in the other, will be
considered too (the sign of the interaction may be selectively reversed by
means of the Feshbach resonance \cite{Feshbach}).

The basic model that will be dealt with in this work amounts to the
following system of linearly coupled normalized GPEs for the mean-field BEC\
wave functions in the two cores, $\psi (x,t)$ and $\phi (x,t)$:
\begin{eqnarray}
i\psi _{t}+\psi _{xx}+\varepsilon \cos (2x)\psi +g_{1}|\psi |^{2}\psi
+\kappa \phi &=&0  \notag \\
&&  \label{model_1d} \\
i\phi _{t}+\phi _{xx}+\varepsilon \cos (2x+\Delta )\phi +g_{2}|\phi
|^{2}\phi +\kappa \psi &=&0  \notag
\end{eqnarray}%
where $\varepsilon $ is the strength of the OL, $g_{1,2}=\pm 1$ are signs of
the nonlinearity $\kappa $ is the linear-coupling coefficient, accounted for
by the linear tunneling between the cores, and $\Delta $ takes into regard a
possible phase shift (\textit{mismatch}) between the OLs in the coupled
cores. The above-mentioned symmetry breaking takes place in the system with $%
\Delta =0$; a change of the symmetry-breaking bifurcations under the action
of the mismatch will be considered too.

In the limit of a very deep OL ($\varepsilon \rightarrow \infty
$), arguments similar to those applied to the single GPE
\cite{Smerzi} suggest that the symmetric version of Eqs.
(\ref{model_1d}) reduces to a system of two linearly coupled
discrete NLS equations. In fact, a system of that type was
introduced in Ref. \cite{DeAngelis}, as a model of a
photonic-crystal coupler for optical signals.

The use of Eqs. (\ref{model_1d}) in the effectively one-dimensional (1D)
form implies that both cores are subjected to tight transverse confinement,
thus allowing one to reduce the underlying three-dimensional GPE to its
counterparts in 1D, as elaborated in several works \cite{3D-1D}. Then, the
temporal and spatial variables in Eqs. (\ref{model_1d}) are related to
physical time $T$ and axial coordinate $X$ by
\begin{equation}
t\equiv T\left( \pi ^{2}\hslash /md^{2}\right) ,~x\equiv \sqrt{2}\pi X/d,
\label{scaling}
\end{equation}%
where $m$ is the atomic mass, and $d$ the OL period in physical units.
Further, the scaled 1D wave functions are related to their 3D counterparts, $%
\Psi $ and $\Phi $, as follows:
\begin{equation}
\left\{
\begin{array}{c}
\Psi (X,R,T) \\
\Phi (X,R,T)%
\end{array}%
\right\} =e^{-i\omega _{\perp }T}\sqrt{\frac{\pi }{2\left\vert
a_{s}\right\vert d^{2}}}\left\{
\begin{array}{c}
\psi (x,t) \\
\phi (x,t)%
\end{array}%
\right\} \exp \left( -\frac{\omega _{\perp }m}{2\hbar }R^{2}\right) ,
\label{psi}
\end{equation}%
where $\omega _{\perp }$ and $R$ are the transverse trapping frequency and
radial coordinate (around each core), and $a_{s}$ the $s$-wave scattering
length of atomic collisions (it is negative if the interaction is
attractive).\textrm{\ }Due to the normalizations, the lattice strength is
represented by $\varepsilon =E_{0}/E_{\mathrm{rec}}$, where $E_{\mathrm{rec}%
}=\left( \pi \hslash \right) ^{2}/\left( md^{2}\right) $\ is the lattice
recoil energy, and $E_{0}$ is the depth of the periodic potential in
physical units.

For the experimental realization of the AA and RR systems, respectively, $%
^{7}$Li \cite{Lithium} and $^{87}$Rb \cite{Markus} condensates  are
appropriate. Results for solitons presented below are in the ballpark of $%
\kappa \sim 1$ in Eqs. (\ref{model_1d}). With regard to Eqs. (\ref{scaling}%
), the corresponding tunnel-coupling time, which is $t_{\mathrm{coupl}}=\pi
/(2\kappa )$ in normalized units, translates, for $d=1$ $\mu $m, into
physical coupling time $T_{\mathrm{coupl}}\sim 10$ $\mu $s and $100$ $\mu $%
s, for lithium and rubidium, respectively. As for the number of atoms in the
solitons, in the AA system is scales, in the normalized units, between $%
N\simeq 2$ and $N\simeq 15$, see Fig. (\ref{solitons1d_11}) below.
As follows from Eqs. (\ref{scaling}) and (\ref{psi}), this
corresponds to the actual number of $^{7}$Li atoms per soliton
between $10^{4}$ and $10^{5}$, if experimentally relevant values
are assumed, $\omega _{\perp }\simeq 2\pi \times 1$ KHz and
$a_{s}\simeq -0.15$ nm (in the experiments without the OL, the
number of atoms in the solitons was up to $5,000$ \cite{Lithium}).
Similarly, in the RR model the number of $^{87}$Rb atoms in gap
solitons is expected to be $\sim 1000$ (in the first experiments,
it was $\simeq 250$ \cite{Markus}).

Equations (\ref{model_1d}) may also be interpreted in terms of nonlinear
optics, describing spatial distribution of light in two parallel planar
waveguides, with periodic transverse modulation of the refractive index,
i.e., as a linear-coupled pair of waveguides considered in Ref. \cite{Wang}.
In that case, $t$ is the propagation distance, $x\ $is the transverse
coordinate, while $\psi $ and $\phi $ are amplitudes of the electromagnetic
fields in the planar cores.

Besides the model based on Eqs. (\ref{model_1d}), we will also use a
modified one, which additionally includes nonlinear coupling between $\psi $
and $\phi $, see Eqs. (\ref{binary}) below. It pertains to a single-core
trap equipped with the OL, that contains a \textit{BEC\ mixture} of two
different spin states of the same atom \cite{mixture}. The linear-coupling
term represents the linear interconversion between the states induced by a
spin-flipping electromagnetic wave \cite{flip}. While various effects were
predicted in the latter setting \cite{mixed}, the spontaneous symmetry
breaking in two-component solitons, that we consider in this work, was not
studied before.

The paper is organized as follows. In Section II, we analyze the linear
spectrum of the coupled system. Methods used for the analysis of soliton
solutions are summarized in Section III. Section IV reports results for
fundamental (even) and twisted (odd) solitons, as well as bound states of
solitons, in symmetric systems, of both the AA and RR types. Results for
asymmetric systems (of the RA type, as well as in systems with mismatched
OLs) are collected in Section V, including a newly found \textit{%
pseudo-bifurcation}, in the system with $\Delta =\pi $. A summary of results
obtained in the model which includes nonlinear interaction between the
coupled components is given in Section VI, and Section VII\ concludes the
paper.

\section{Linear spectrum}

\begin{figure}[tbp]
\subfigure[]{\includegraphics[width=3in]{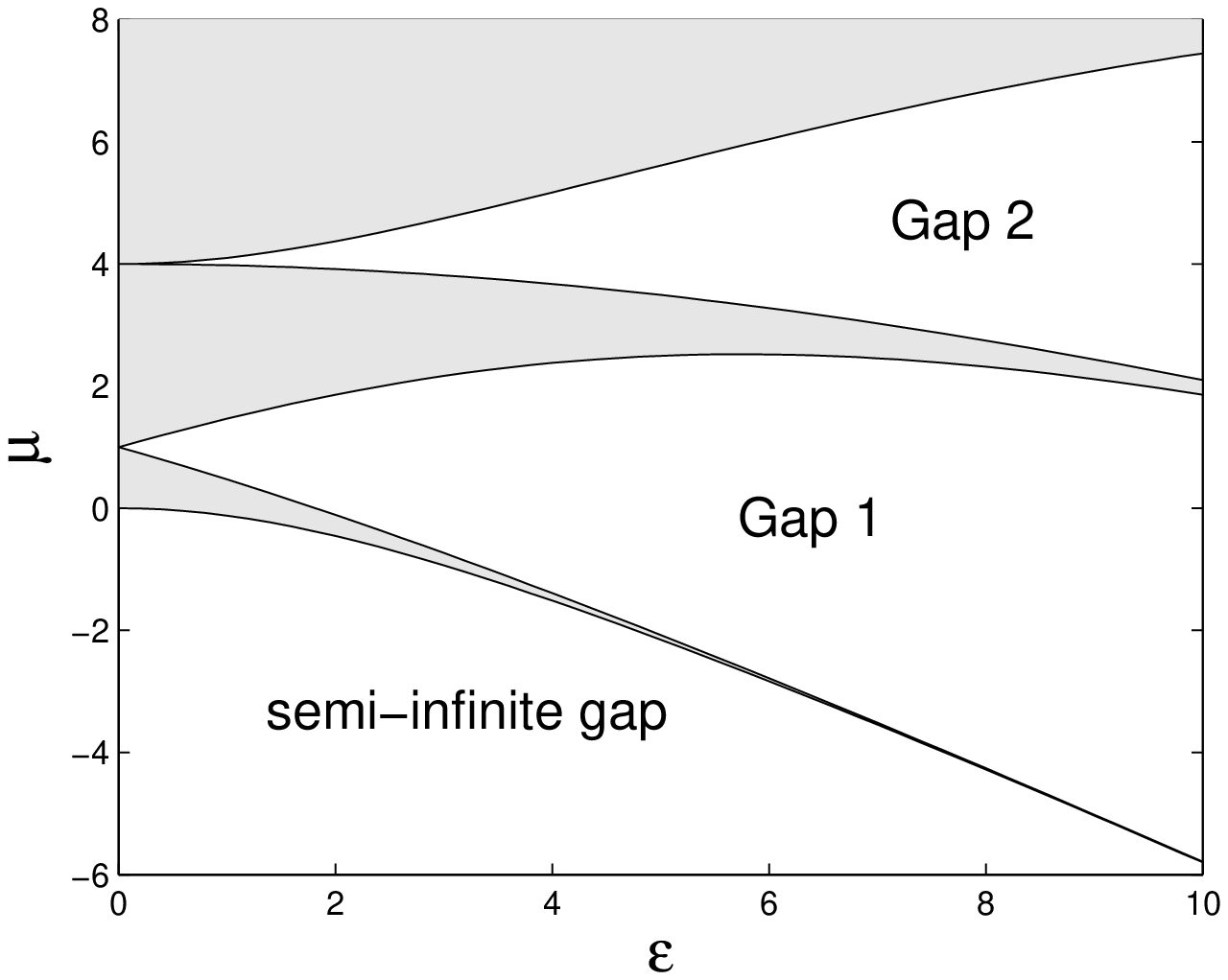}} \subfigure[]{%
\includegraphics[width=3in]{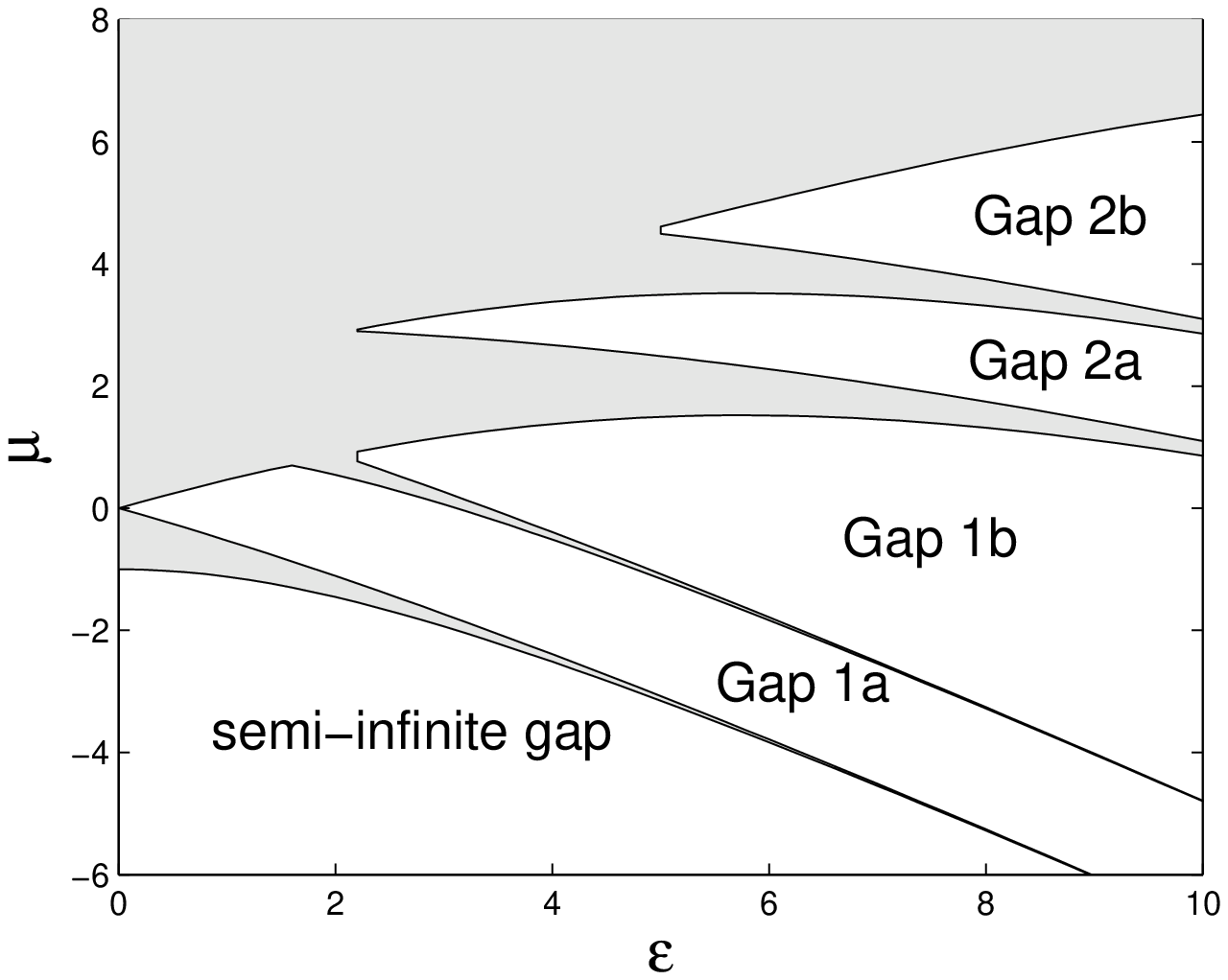}}
\caption{Transformation of the linear spectrum with the increase of the
lattice strength $\protect\varepsilon $. Here and in other figures including
the spectrum, the shaded areas show Bloch bands. (a) The ordinary
(decoupled) Mathieu equation, i.e., each equation (\protect\ref{mu1}) and (%
\protect\ref{mu2}) with $\protect\kappa =0;$ (b) the coupled system, with $%
\protect\kappa =1$ and $\Delta =0$.}
\label{spectrum_fig}
\end{figure}

Our first objective is to examine how the linear coupling, accounted for by
coefficient $\kappa $ in Eqs. (\ref{model_1d}), alters the spectrum of the
linear system (first, in the model with aligned lattices, $\Delta =0$). We
look for solutions to the linearized equations as $\left\{ \psi (x,t),\phi
(x,t)\right\} =\left[ \alpha (x)\pm \beta (x)\right] e^{-i\mu t}$, with
chemical potential $\mu $. This leads to decoupled Mathieu equations (ME)
with eigenvalues $\mu \pm \kappa $,%
\begin{eqnarray}
\alpha ^{\prime \prime }+\varepsilon \cos (2x)\alpha (x)+(\mu +\kappa
)\alpha (x) &=&0,  \label{mu1} \\
\beta ^{\prime \prime }+\varepsilon \cos (2x)\beta (x)+(\mu -\kappa )\beta
(x) &=&0,  \label{mu2}
\end{eqnarray}%
where the prime stands for $d/dx$. The ME spectrum gives rise to the
well-known bandgap structure. Note that, in the model of two linearly
coupled Bragg gratings (i.e., a system of four equations for
counterpropagating waves in two cores), the coupling leads to shrinkage or
closure of the spectral gap \cite{Mak}. In the present case, the effect of
the coupling is more complex. Given $\mu $ belongs to a bandgap in the
present model if $\mu $ falls into one of the gaps in \emph{both} equations (%
\ref{mu1}) and (\ref{mu2}). Further consideration demonstrates that,
depending on the values of $\kappa $ and OL strength $\varepsilon $, each
gap originating from the ME spectrum either shrinks (or, sometimes,
completely closes up), similar to the situation in the above-mentioned model
of linearly coupled Bragg gratings, or splits into pairs of \textit{subgaps}%
. An example of the splitting of the first two gaps [Gap \{1,2\} $%
\rightarrow $ Gaps \{(1a,1b),(2a,2b)\}] under the action of the linear
coupling is displayed in Fig. \ref{spectrum_fig}.

Unlike Eqs. (\ref{mu1}) and (\ref{mu2}), the linearized equations do not
decouple with $\Delta \neq 0$ in Eqs. (\ref{model_1d}). A typical example of
the dependence of the spectrum on $\Delta $ is shown in Fig. \ref%
{spectrum_delta}. With the increase of $\Delta $, the subgaps (that have
appeared as a result of the coupling-induced splitting) shrink. At $\Delta \
$close to $\pi /2$, gap 2a disappears, which is followed by the
disappearance of gap 1a. At $\Delta =\pi $, the entire spectrum
qualitatively resembles that of the ordinary ME.

\begin{figure}[tbp]
\centering\includegraphics[width=3in]{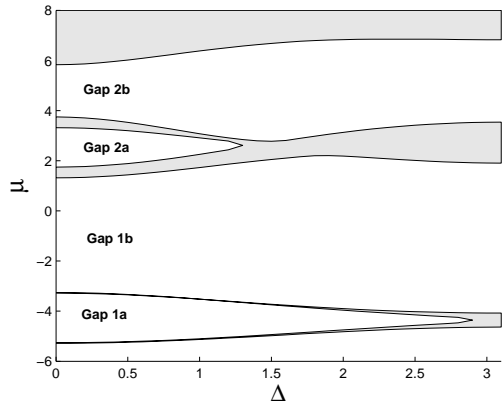}
\caption{The transformation of the linear spectrum with the increase of
mismatch $\Delta $ for $\protect\varepsilon =8$, $\protect\kappa =1$.}
\label{spectrum_delta}
\end{figure}

\section{Analysis of soliton solutions}

\subsection{Symmetric, antisymmetric, and asymmetric solitons}

Stationary solutions to Eqs. (\ref{model_1d}) are looked as $\left\{ \psi
,\phi \right\} =\left\{ u(x),v(x)\right\} e^{-i\mu t}$, with \text{real}
functions $u$ and $v$ obeying the equations%
\begin{equation}
\begin{array}{c}
\mu u+u^{\prime \prime }+\varepsilon \cos (2x)u+g_{1}u^{3}+\kappa v=0, \\
\mu v+v^{\prime \prime }+\varepsilon \cos (2x+\Delta )v+g_{2}v^{3}+\kappa
u=0.%
\end{array}
\label{uv}
\end{equation}%
First, we explore the symmetries of the system in the AA ($g_{1}=g_{2}=+1$)
and RR ($g_{1}=g_{2}=-1$) cases with aligned OLs, $\Delta =0$. To this end,
we designate $\hat{u}_{0}(x;\mu )$ the solution of the single-component GPE,
i.e. either of two equations (\ref{uv}) with $\kappa =\Delta =0$,
corresponding to chemical potential $\mu $. Then, Eqs. (\ref{uv}) give rise
to symmetric and antisymmetric solitons,%
\begin{equation}
u(x;\mu )=v(x;\mu )=\hat{u}_{0}(x;\mu +\kappa ),  \label{symm_sol_1d}
\end{equation}%
\begin{equation}
u(x;\mu )=-v(x;\mu )=\hat{u}_{0}(x;\mu -\kappa ).  \label{antisymm_sol_1d}
\end{equation}%
If the linear coupling splits the bandgap into two subgaps as outlined
above, the shifts of the chemical potential as per Eqs. (\ref{symm_sol_1d})
and (\ref{antisymm_sol_1d}), $\mu \rightarrow \mu \pm \kappa $, displace the
symmetric and antisymmetric soliton to the lower and higher subgap,
respectively. In particular, in the coupled AA system, symmetric solitons,
which are generated by their counterparts belonging to the semi-infinite gap
in the single-component GPE with attraction according to Eq. (\ref%
{symm_sol_1d}), remain in the semi-infinite gap, while the corresponding
anti-symmetric solitons may either stay in the semi-infinite gap, or move to
subgap 1a (in the case shown in Fig. \ref{spectrum_fig}), or even end up
inside the Bloch band separating the semi-infinite gap and subgap 1a. In the
latter case, the antisymmetric soliton becomes \textit{embedded} (into the
continuous spectrum), as defined in Refs. \cite{embedded}.

The central issue to be considered below is a possibility of \textit{%
spontaneous symmetry breaking}, i.e. finding a bifurcation point beyond
which asymmetric solitons emerge. To quantify the symmetry breaking, we
define norms in the two cores, and the asymmetry ratio, $\Theta $, as%
\begin{equation}
N_{u,v}\equiv \int_{-\infty }^{+\infty }\left\{ u^{2}(x),v^{2}(x)\right\}
dx,~\Theta \equiv \frac{\left\vert N_{u}-N_{v}\right\vert }{N_{u}+N_{v}}
\label{theta}
\end{equation}%
[Eqs. (\ref{model_1d}) conserve only the total norm, $N\equiv N_{u}+N_{v}$].
The bifurcation was found from a numerical solution of Eqs. (\ref{uv}). The
results are reported in the next section for the models of the AA, RR, and
RA types.

\subsection{Soliton stability}

To tackle the stability problem, perturbed solutions are taken as%
\begin{eqnarray}
\left\{ \psi (x,t)\,,\phi (x,t)\right\} &=&\left\{ \left[ \hat{u}(x;\mu
)+\xi _{1}(x)e^{i\eta t}\right] ,\left[ \hat{v}(x;\mu )+\xi _{3}(x)e^{i\eta
t}\right] \right\} e^{-i\mu t},  \notag \\
&&  \label{perturbed_1d} \\
\left\{ \psi ^{\ast }(x,t),\phi ^{\ast }(x,t)\right\} &=&\left\{ \left[ \hat{%
u}^{\ast }(x;\mu )+\xi _{2}(x)e^{i\eta t}\right] ,\left[ \hat{v}^{\ast
}(x;\mu )+\xi _{4}(x)e^{i\eta t}\right] \right\} e^{i\mu t},  \notag
\end{eqnarray}%
where $\left\{ \hat{u}(x;\mu ),\hat{v}(x;\mu )\right\} e^{-i\mu t}$, with
real functions $\hat{u}(x;\mu )$ are $\hat{v}(x;\mu )$, are stationary
solutions of Eqs. (\ref{uv}). Perturbations $\xi _{1,3}$ of fields $\psi
,\phi $, and their complex conjugates, $\xi _{2,4}$, are formally treated as
independent functions. Substituting expressions (\ref{perturbed_1d}) in Eqs.
(\ref{model_1d}) and linearizing, we arrive at an eigenvalue problem,%
\begin{equation}
\begin{bmatrix}
\mathcal{L}\left( \hat{u}(x;\mu ),\mu ,g_{1},0\right) & \kappa \sigma _{3}
\\
\kappa \sigma _{3} & \mathcal{L}\left( \hat{v}(x;\mu ),\mu ,g_{2},\Delta
\right)%
\end{bmatrix}%
\begin{bmatrix}
\xi _{12} \\
\xi _{34}%
\end{bmatrix}%
=\eta
\begin{bmatrix}
\xi _{12} \\
\xi _{34}%
\end{bmatrix}%
,  \label{full_eig_1d}
\end{equation}%
where $\xi _{12}\equiv \left( \xi _{1},\xi _{2}\right) ^{T}$, $\xi
_{34}\equiv \left( \xi _{3},\xi _{4}\right) ^{T}$, the linear-stability
operator for the single-component GPE is
\begin{equation}
\mathcal{L}\left( f(x),\mu ,g,\Delta \right) =\left[ \frac{d^{2}}{dx^{2}}%
+\varepsilon \cos (2x+\Delta )+2gf^{2}(x)+\mu \right] \sigma
_{3}+igf^{2}(x)\sigma _{2},
\end{equation}%
and $\sigma _{3}$ and $\sigma _{2}$ are the Pauli matrices. In the
single-component GPE, families of fundamental solitons have been shown to be
stable \cite{Pelinovsky}, i.e., all eigenvalues of the corresponding
operator $\mathcal{L}\left( \hat{u}_{0}(x;\mu ),\mu ,g,0\right) $ are real.

Equation (\ref{full_eig_1d}) can be simplified for symmetric and
anti-symmetric solutions. Substituting, respectively, expressions (\ref%
{symm_sol_1d}) and (\ref{antisymm_sol_1d}) (with $\Delta =0$ and $%
g_{1}=g_{2}\equiv g$), we arrive at a reduced eigenvalue problem for the
symmetric soliton,%
\begin{eqnarray}
\mathcal{L}\left( \hat{u}_{0}(x;\mu +\kappa ),\mu +\kappa ,g,0\right) \xi
_{+} &=&\eta \xi _{+},  \label{symm_reduced_a} \\
\mathcal{L}\left( \hat{u}_{0}(x;\mu +\kappa ),\mu -\kappa ,g,0\right) \xi
_{-} &=&\eta \xi _{-}  \label{symm_reduced_b}
\end{eqnarray}%
($\xi _{\pm }\equiv \xi _{12}\pm \xi _{34}$), and its counterpart for the
anti-symmetric one,%
\begin{eqnarray}
\mathcal{L}\left( \hat{u}_{0}(x;\mu -\kappa ),\mu +\kappa ,g,0\right) \xi
_{+} &=&\eta \xi _{+},  \label{antisymm_reduced_a} \\
\mathcal{L}\left( \hat{u}_{0}(x;\mu -\kappa ),\mu -\kappa ,g,0\right) \xi
_{-} &=&\eta \xi _{-}.  \label{antisymm_reduced_b}
\end{eqnarray}

Since Eqs. (\ref{symm_reduced_a}) and (\ref{antisymm_reduced_b}) are
tantamount to the linear-stability problem in the single-component GPE, the
stability of the soliton in the latter equation is a necessary condition for
the full stability of symmetric and antisymmetric solitons in the coupled
system. To identify the full stability conditions for the symmetric and
antisymmetric solitons, we solved the additional eigenvalue problems, i.e.,
respectively, Eqs. (\ref{symm_reduced_b}) and (\ref{antisymm_reduced_a}).
The stability of asymmetric solitons was inferred from a numerical solution
of the full eigenvalue problem, based on Eq. (\ref{full_eig_1d}). The
solution is stable if all respective eigenvalues $\eta $ are real.

\section{Results (symmetric systems)}

\subsection{The attraction-attraction model}

\begin{figure}[tbp]
\subfigure[]{\includegraphics[width=3in]{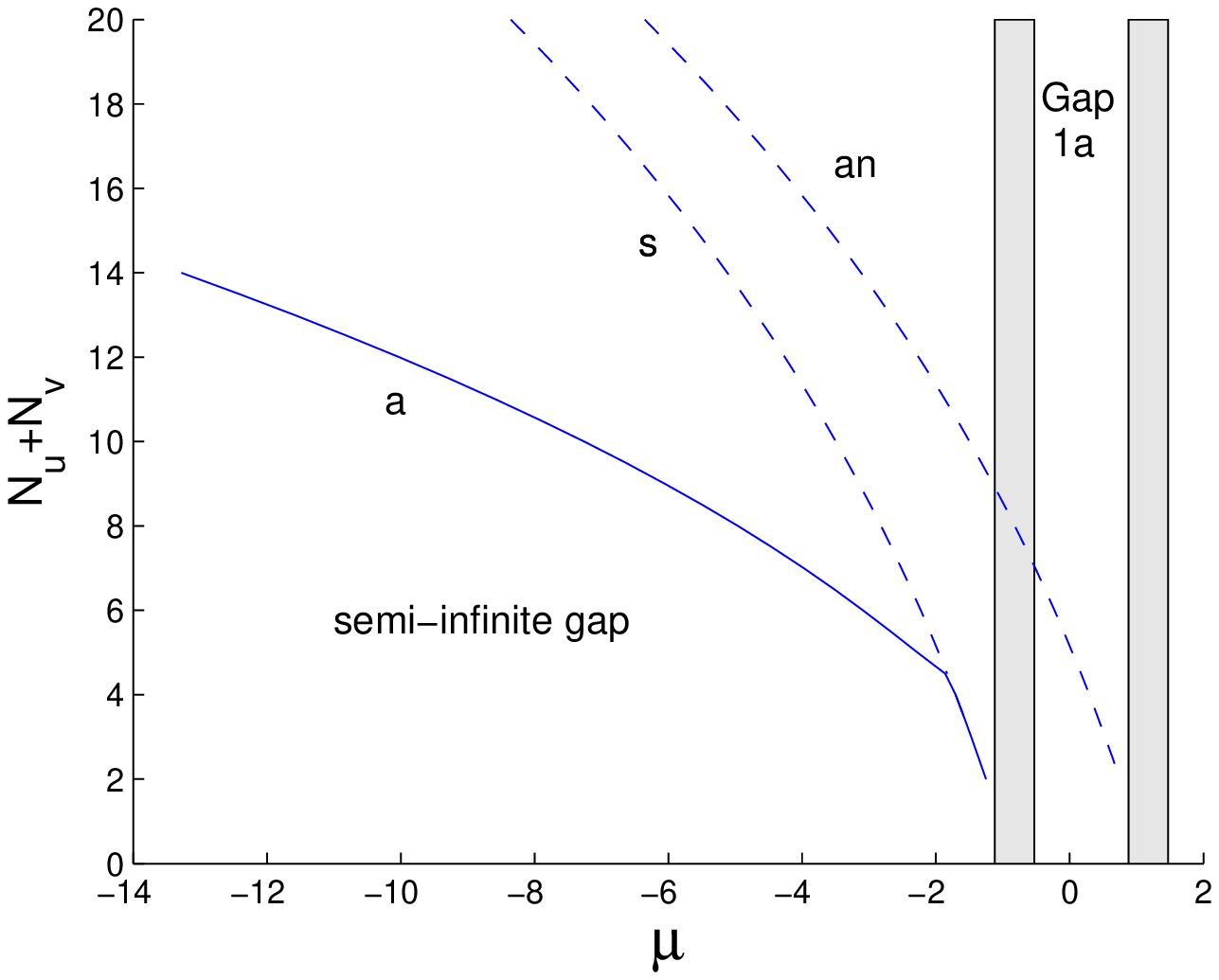}\label{fig2_weak}}  %
\subfigure[]{\includegraphics[width=3in]{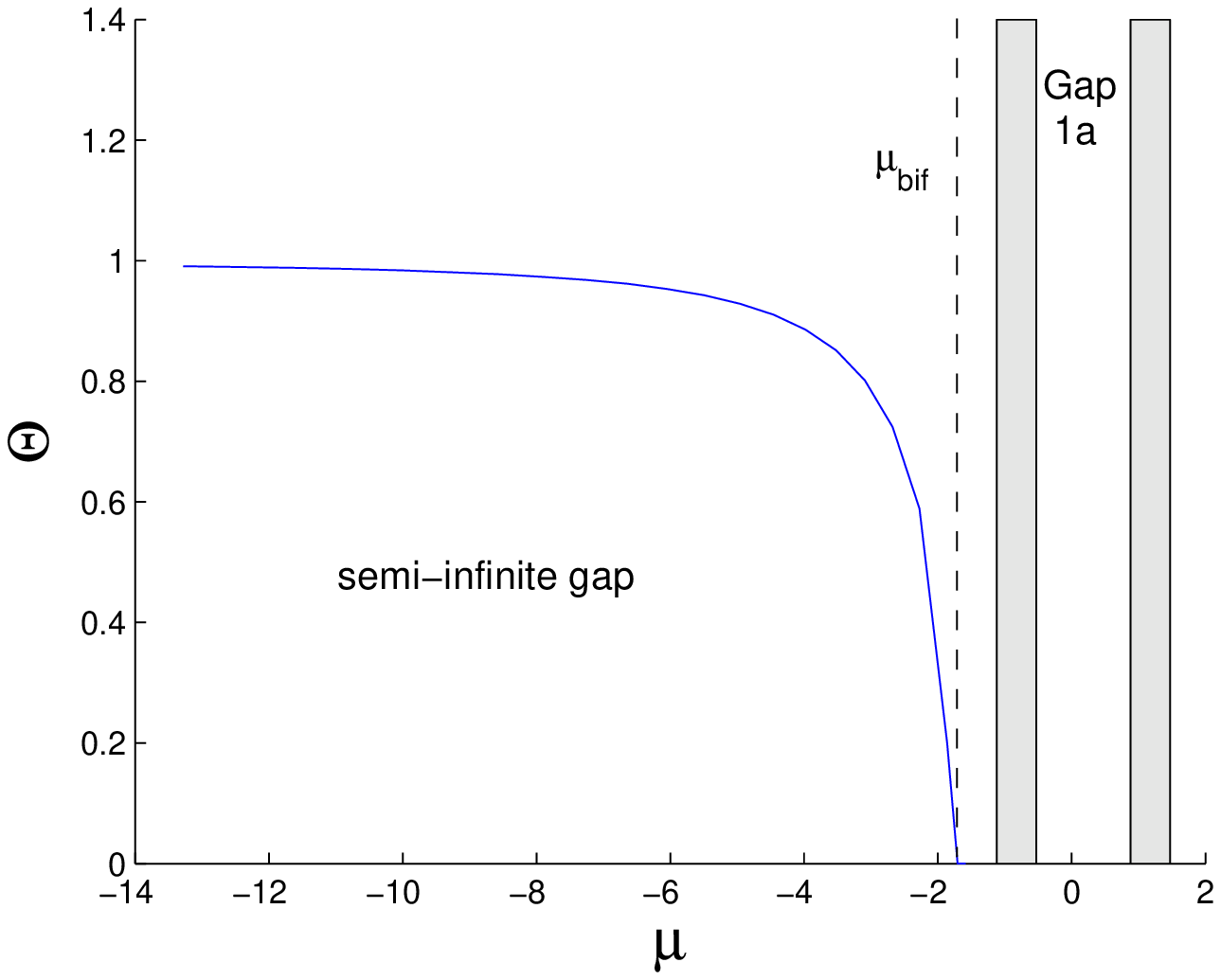}} \subfigure[]{%
\includegraphics[width=3in]{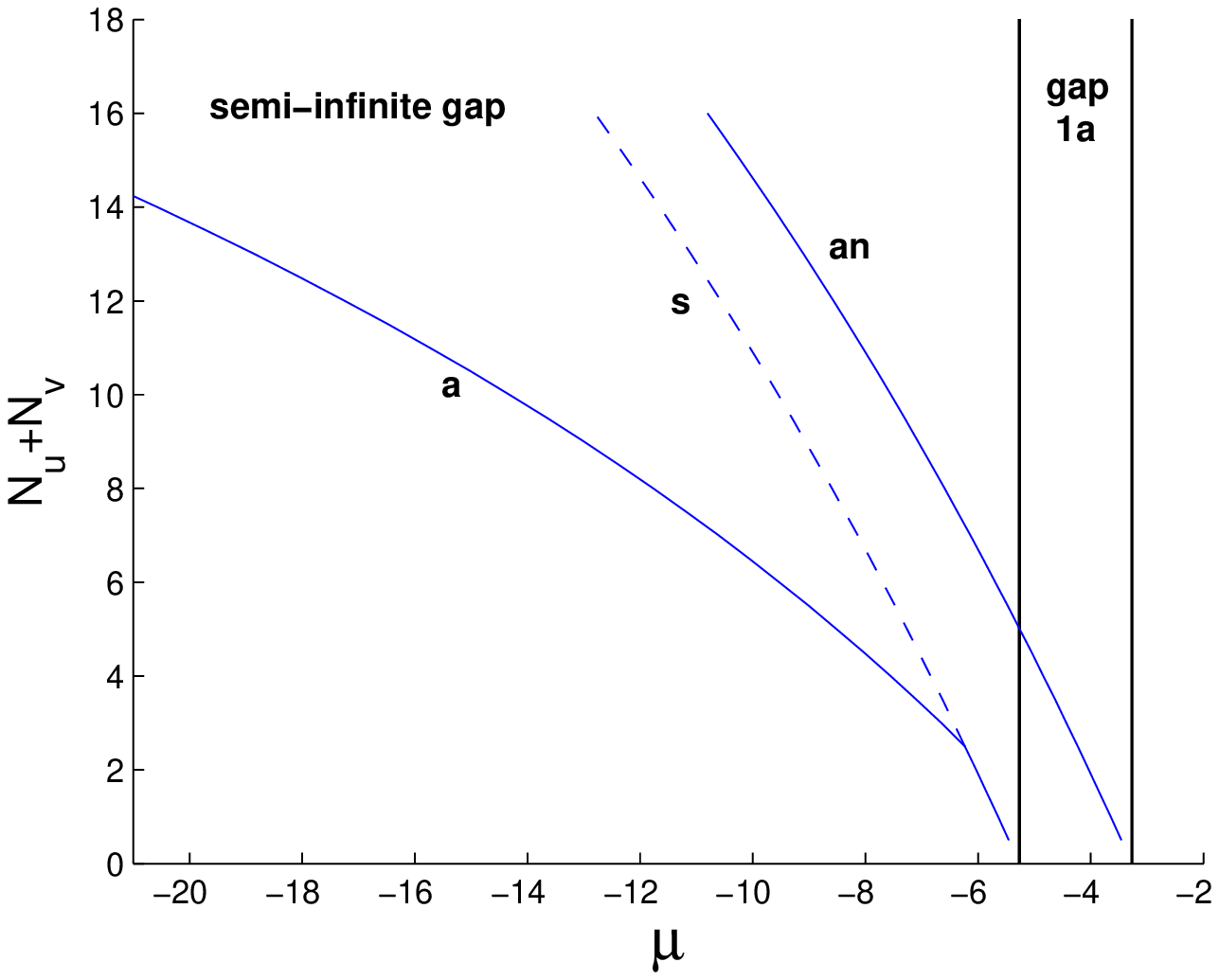}\label{fig2_strong}}  %
\subfigure[]{\includegraphics[width=3in]{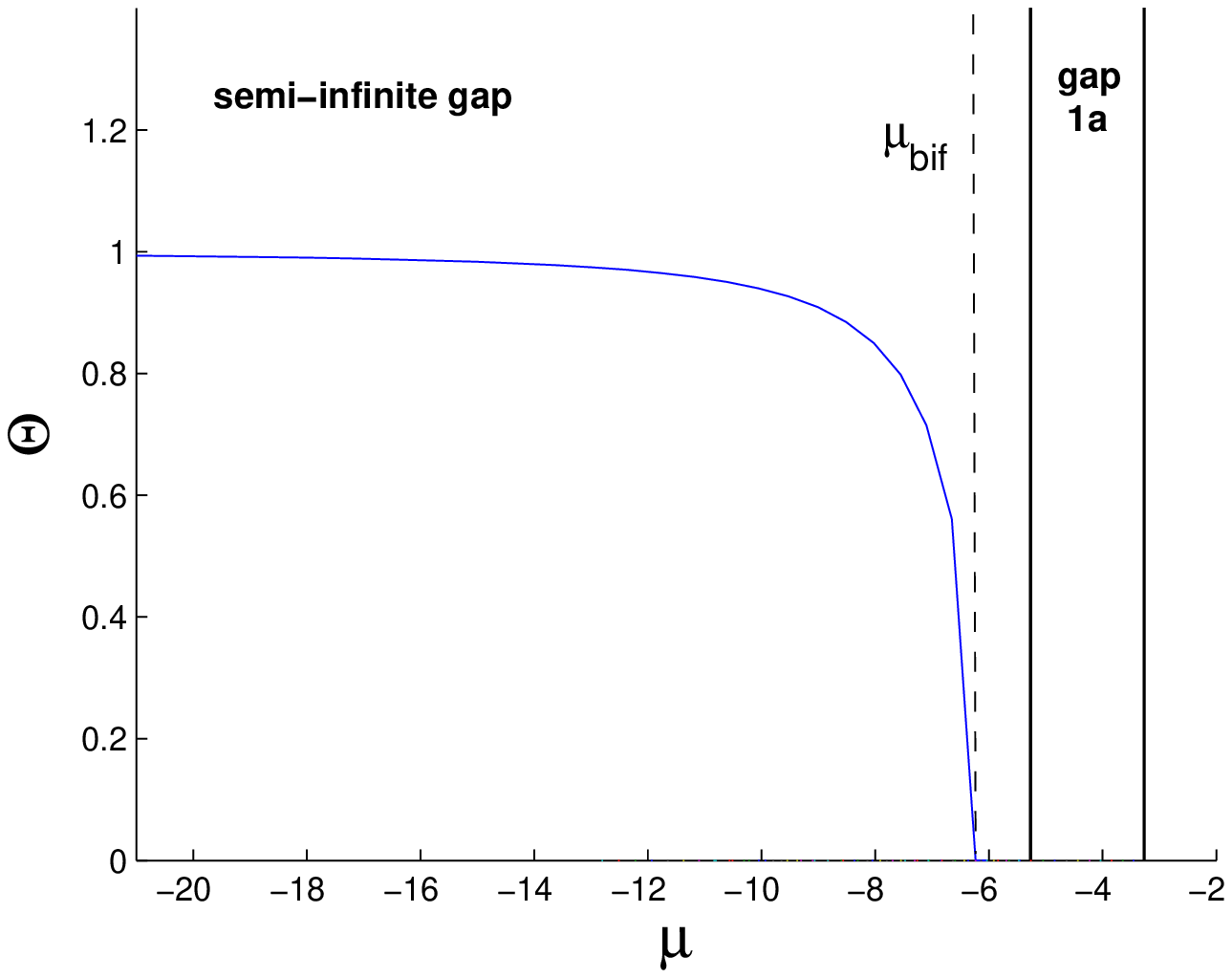}}
\caption{(Color online) Soliton families in the attraction-attraction model
with $\protect\kappa =1$. In this figure and below, the results are
displayed for the dual-core model with aligned lattices ($\Delta =0$),
unless a nonzero value of $\Delta $ is indicated. Presented are cases of
relatively weak, $\protect\varepsilon =1$ (a,b), and strong, $\protect%
\varepsilon =8$ (c,d), lattices. Here and below, stable and
unstable branches are shown by solid and dashed lines, with labels
``s", ``an" and ``a" referring to symmetric, antisymmetric, and
asymmetric solutions, respectively. (a,c): The total norm of the
soliton versus its chemical potential, $\protect\mu $. (b,d): The
asymmetry ratio [defined as per Eq. (\protect\ref{theta})] versus $\protect%
\mu $, for families of asymmetric solitons; the vertical line labeled $%
\protect\mu _{\mathrm{bif}}$ marks the bifurcation point. Note that the
branch of anti-symmetric solutions is completely unstable in the weak
lattice (a,b), and completely stable in the strong one (b,d).}
\label{solitons1d_11}
\end{figure}

Families of numerically found stationary soliton solutions of Eqs. (\ref{uv}%
) for the AA system ($g_{1}=g_{2}=+1$) with zero mismatch ($\Delta =0$) are
displayed in Fig. \ref{solitons1d_11} for $\kappa =1$, in the cases of weak (%
$\varepsilon =1$) and strong ($\varepsilon =8$) OLs. We observe that a
branch of asymmetric solitons bifurcates from the symmetric one, while the
antisymmetric solutions do not give rise to any bifurcation. We also notice
that the antisymmetric branch extends through the Bloch band separating the
semi-infinite gap and the first finite bandgap, 1a; as said above, the
solitons are \textit{embedded} ones inside the band. Panels (b) and (d) in
the figure clearly show that the symmetry-breaking bifurcation is\textit{\
supercritical}\ (the bifurcating branch immediately goes \emph{forward} in $%
-\mu $, without turning backward, the latter being a
characteristic feature of the \textit{subcritical} bifurcation --
in particular, in the system of linearly coupled NLS equations of
the AA type without the lattice potential \cite{dual-core}). The
present situation is, generally, similar to that in the system of
linearly coupled Bragg gratings \cite{Mak} (where the bifurcation
is supercritical too).

\begin{figure}[tbp]
\subfigure[]{\includegraphics[width=3in]{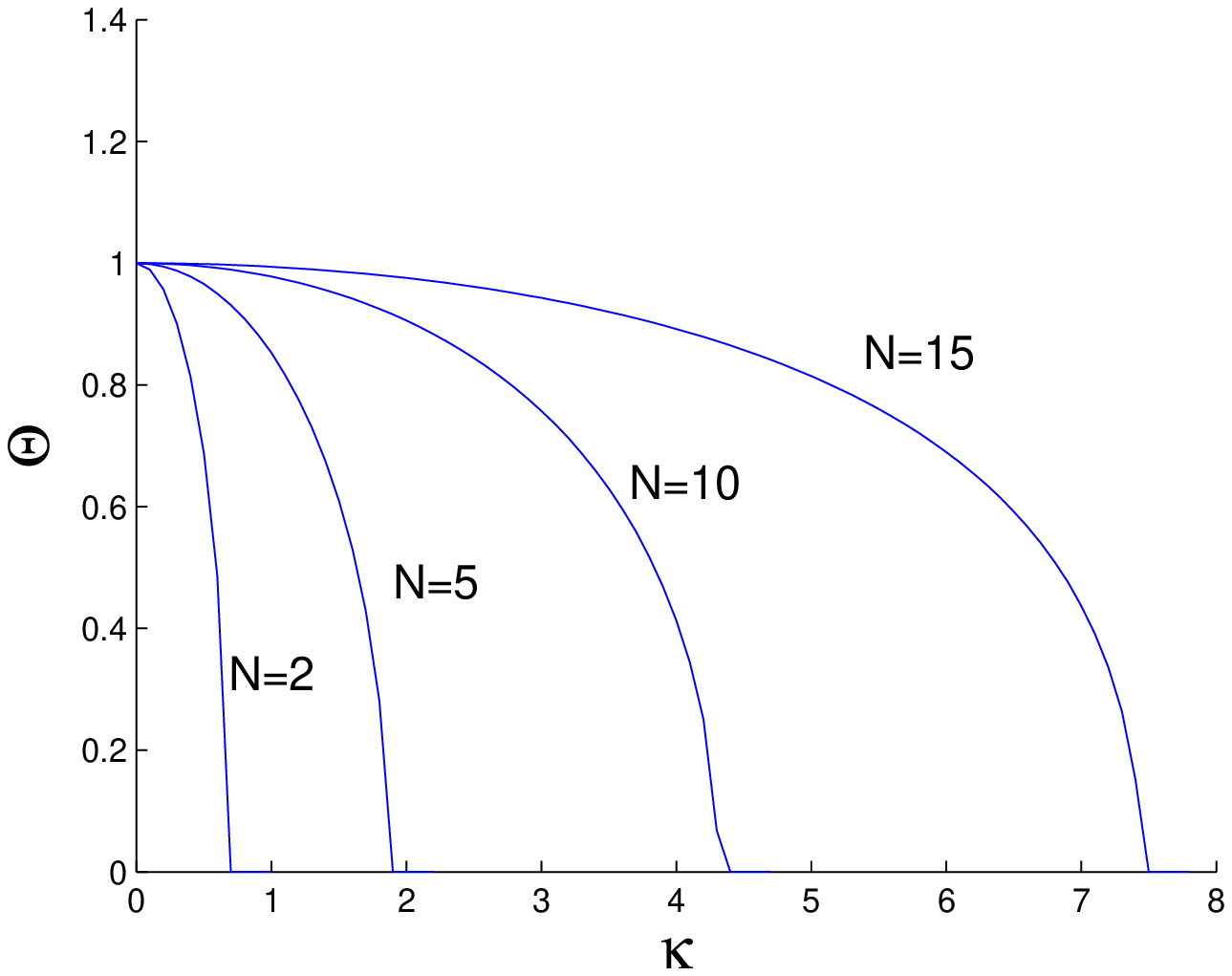}} \subfigure[]{%
\includegraphics[width=3in]{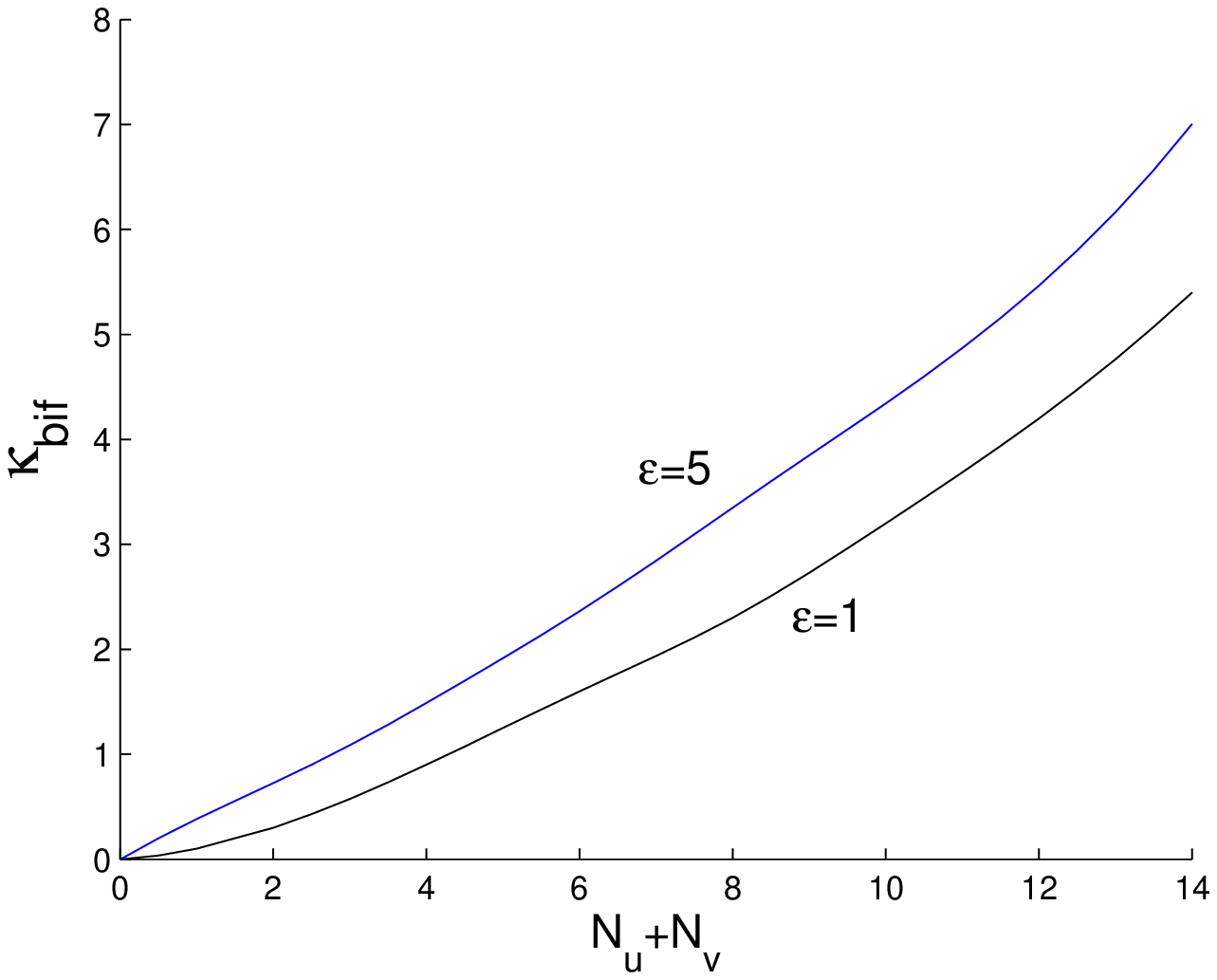}}
\caption{(Color online) Properties of asymmetric solitons in the
attraction-attraction system. (a) The asymmetry ratio [defined as per Eq. (%
\protect\ref{theta})] as a function of coupling coefficient $\protect\kappa $
at $\protect\varepsilon =5$ and different values of the soliton's norm. (b)
The value of $\protect\kappa $ at the bifurcation point versus the soliton's
norm at different values of the lattice strength, $\protect\varepsilon $.}
\label{solitons1d_11_properties}
\end{figure}

The change of the bifurcation with variation of coupling constant $\kappa $
is illustrated by Fig. \ref{solitons1d_11_properties}. In particular, the$\ $%
critical value of the coupling constant, $\kappa _{\mathrm{bif}}$, at which
the bifurcation occurs is shown in panel (b) as a function of the soliton's
total norm [note that in Fig. \ref{solitons1d_11}(a), which pertains to $%
\kappa =1$, the bifurcation in the system with $\varepsilon =1$ occurs at $%
N\approx 4$, in compliance with Fig. \ref{solitons1d_11_properties}(b)].
Naturally, $\kappa _{\mathrm{bif}}$ grows with $N$, as the bifurcation is a
result of the competition between the nonlinear and linear properties of the
system, accounted for by $N$ and $\kappa $, respectively. On the other hand,
the dependence of the bifurcation point on the lattice's depth $\varepsilon $
is quite weak, the stronger lattice making the region of the existence of
asymmetric solitons somewhat larger (which can be easily understood too, as
the lattice tends to pin and thus stabilize any localized pattern).%

The above results bear some implications for stability of the solutions, as,
in models of the NLS type with self-focusing, a necessary stability
condition is given by the Vakhitov-Kolokolov (VK)\ criterion \cite{VK}. In
the present notation, it is $dN/d\mu <0$. As seen in Fig. \ref{solitons1d_11}%
, all solution branches satisfy this condition (however, as shown below, not
all of them are stable, as the VK criterion is not sufficient for the
stability).

\begin{figure}[tbp]
\includegraphics[width=3in]{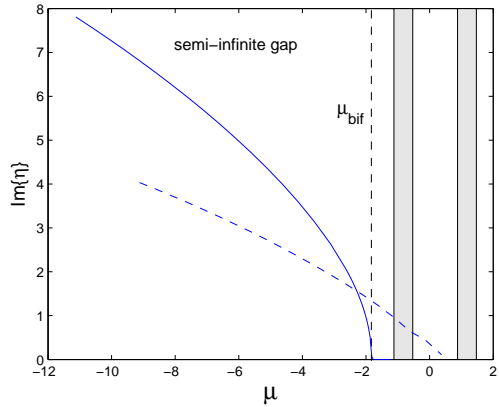}
\caption{(Color online) The largest instability growth rates for symmetric
and anti-symmetric solitons in the attraction-attraction system, with $%
\protect\varepsilon =1$, $\protect\kappa =1$, are shown by the continuous
and dashed curves, respectively.}
\label{solitons1d_11_stability}
\end{figure}

To investigate the stability of the symmetric and antisymmetric soliton
families presented in Fig. \ref{solitons1d_11} in an accurate form, we
solved the reduced eigenvalue problem, based on Eq. (\ref{symm_reduced_b})
or (\ref{antisymm_reduced_a}), respectively, to identify the eigenvalue with
the largest imaginary part, $\mathrm{Im}\{\eta \}$, i.e., the fastest
growing unstable mode. Figure \ref{solitons1d_11_stability} shows the result
for a relatively weak OL, with $\varepsilon =1$. We conclude that the
symmetric soliton is stable up to the bifurcation point, where the branch of
the asymmetric solitons emerges. Beyond this point, the symmetric solitons
are unstable. On the other hand, the branch of the anti-symmetric solitons,
which undergoes no bifurcation in the AA system, is unstable in weak OLs,
see Figs. \ref{fig2_weak} and \ref{solitons1d_11_stability}, but becomes
stable in a stronger lattice, see Fig. \ref{fig2_strong}. In fact, strong
OLs give rise to \textit{bistability}: the antisymmetric soliton coexists,
as a stable solution, with either symmetric or asymmetric one. Finally,
numerical solution of the full eigenvalue problem, based on Eq. (\ref%
{full_eig_1d}), for the asymmetric solitons demonstrates that they are
stable whenever they exist.

\begin{figure}[tbp]
\includegraphics[width=3in]{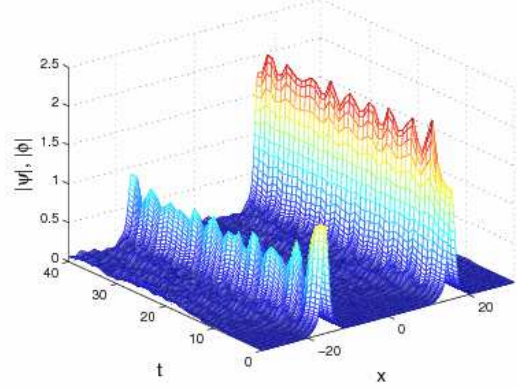}
\caption{(Color online) Perturbation-induced transformation of an unstable
symmetric soliton into a stable asymmetric one in the attraction-attraction
model with a relatively weak lattice ($\protect\varepsilon =1$, $\protect%
\kappa =1$). The initial norm is $N=5.6$.Here and in Fig. \protect\ref%
{odd_evol} below, the two components of the solution are juxtaposed in the
same panel.}
\label{sol1d_11_evolution}
\end{figure}

The predictions of the linear stability analysis have been confirmed by
direct numerical simulations of the underlying equations (\ref{model_1d}).
In particular, Fig. \ref{sol1d_11_evolution} shows that unstable symmetric
solitons evolve into stable asymmetric ones. It has also been checked that
antisymmetric solitons in sufficiently strong OLs are stable indeed, while,
in weak lattices, they are completely destroyed by growing perturbations.

\begin{figure}[tbp]
\centering\includegraphics[width=3in]{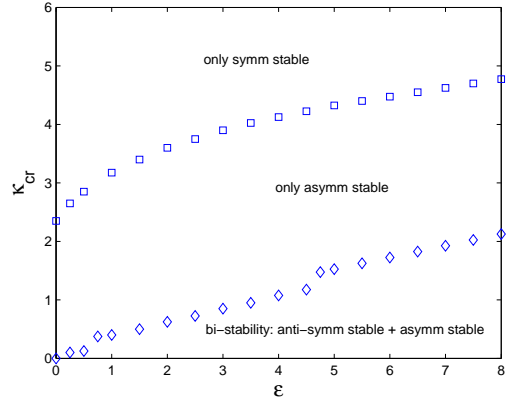}
\caption{(Color online) The attraction-attraction model: critical values of
the coupling coefficient, $\protect\kappa _{\mathrm{cr}}$, versus the
lattice strength $\protect\varepsilon $, for a fixed norm, $N=10$. The upper
curve (squares) is, in fact, the bifurcation line, i.e., the stability
border of the symmetric solitons (which are stable above the border), and
existence border of stable asymmetric solitons (they exist beneath the
border). The lower curve (rhombuses) is the stability border of
antisymmetric solitons, which are stable beneath it, thus making the lower
area a bistability region.}
\label{antisymm_stab_thresh}
\end{figure}

Results of the analysis are summarized in Fig. \ref{antisymm_stab_thresh},
in the form of a diagram showing the stability regions of the symmetric,
antisymmetric, and asymmetric solitons in the $\left( \varepsilon ,\kappa
\right) $ plane. To this end, we define, in addition to the above-mentioned
bifurcation value, $\kappa _{\mathrm{bif}}$, which plays the role of the
instability border of symmetric solitons (they are unstable, for given $%
\varepsilon $ and $N$, at $\kappa <\kappa _{\mathrm{bif}}$, and stable at $%
\kappa \geq \kappa _{\mathrm{bif}}$) and existence border of stable
asymmetric solitons (they exist at $\kappa <\kappa _{\mathrm{bif}}$), a
critical value which separates stable and unstable antisymmetric solitons:
they are stable at $\kappa \leq \kappa _{\mathrm{cr}}$, and unstable at $%
\kappa >\kappa _{\mathrm{cr}}$.

The stability analysis demonstrates that the branch of the antisymmetric
solitons changes its character from unstable to stable \emph{as a whole},
with the increase of $\varepsilon $ at fixed $\kappa $ [see Figs. \ref%
{solitons1d_11}(a) and (c)], or decrease of $\kappa $ at fixed $\varepsilon $%
. In other words, a situation has not been found when a part of the branch
of antisymmetric solitons would be stable, while its other part is unstable.
Note that the segment of the branch extending through the Bloch band, where
the antisymmetric solitons are \textit{embedded} ones, is stable too.

\subsection{The repulsion-repulsion system}

\begin{figure}[tbp]
\subfigure[]{\includegraphics[width=3in]{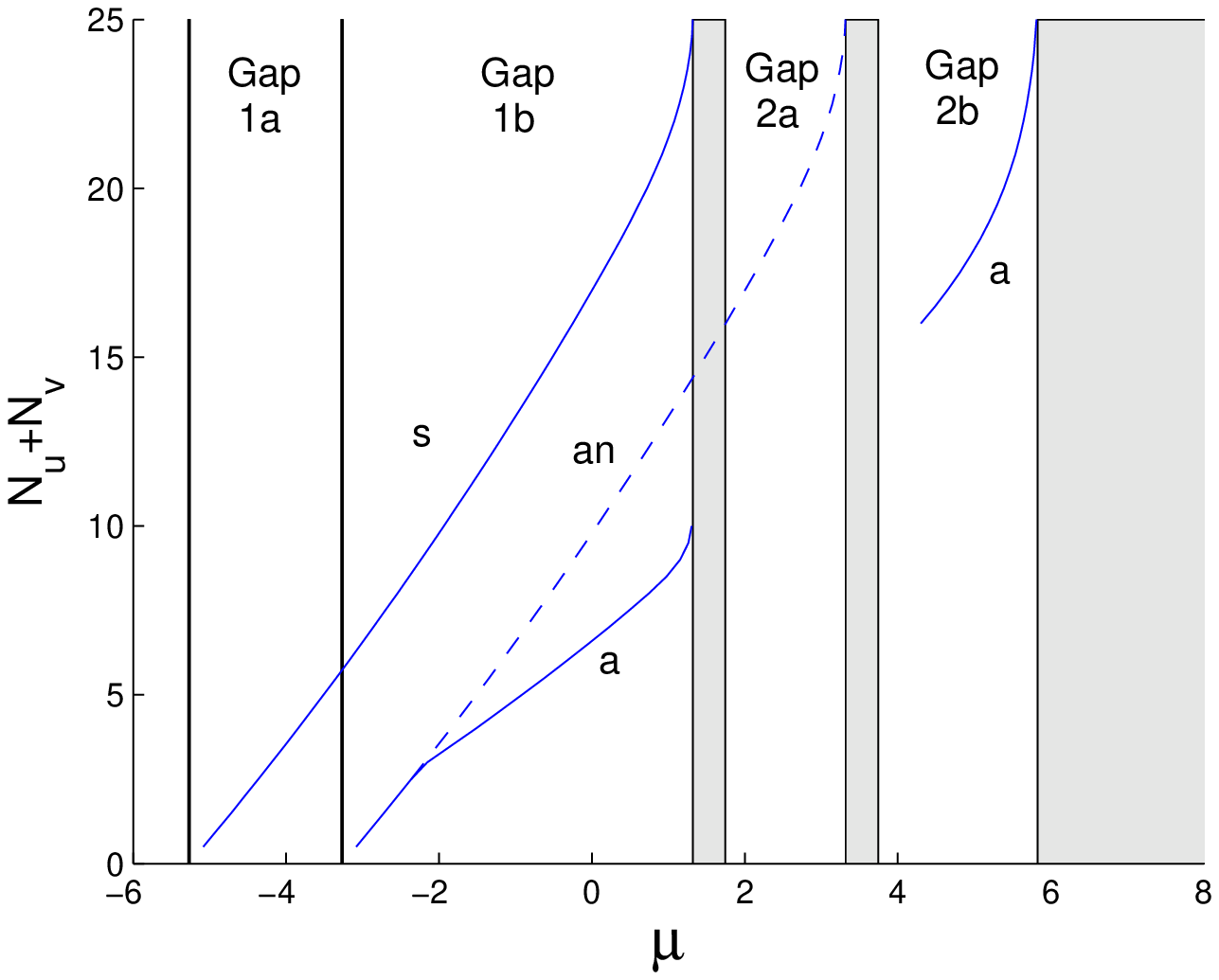}} \subfigure[]{%
\includegraphics[width=3in]{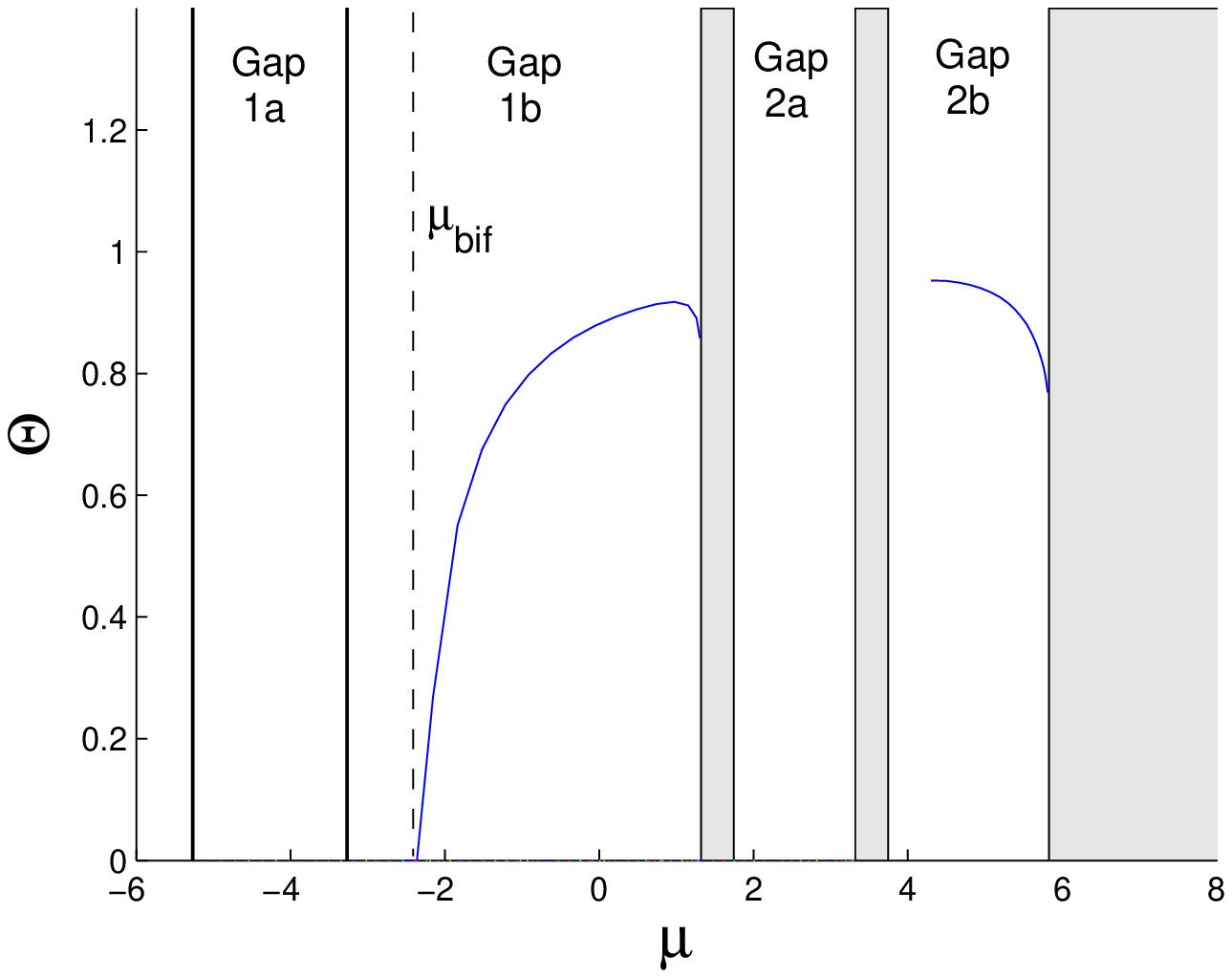}}
\caption{(color online) Soliton families in the repulsion-repulsion model
with $\protect\varepsilon =8$ and $\protect\kappa =1$. Panels (a) and (b)
have the same meaning as in Fig. \protect\ref{solitons1d_11}.}
\label{solitons1d_00a}
\end{figure}

\begin{figure}[tbp]
\includegraphics[width=3in]{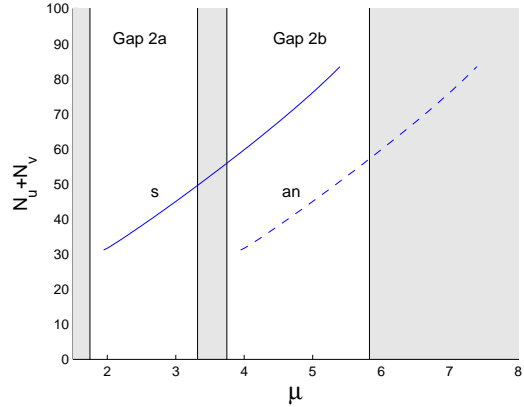}
\caption{(Color online) Continuation of Fig. \protect\ref{solitons1d_00a} to
larger values of the soliton's norm.}
\label{solitons1d_00b}
\end{figure}

In the single-component GPE with repulsion, solitons may only be found in
finite bandgaps, therefore they are called gap solitons \cite{GS,Markus}. In
the coupled system with repulsion in both cores, symmetric and antisymmetric
states may be composed of single-component gap solitons as per Eqs. (\ref%
{symm_sol_1d}) and (\ref{antisymm_sol_1d}). Families of the solitons in the
RR system, generated from the gap solitons belonging to the first and second
finite bandgaps in the single-component GPE, are shown in Figs. \ref%
{solitons1d_00a} and \ref{solitons1d_00b}, along with families
of asymmetric solitons bifurcating from them. In particular, the symmetric
branch in subgaps 2a and 2b in Fig. \ref{solitons1d_00b} (which extends
across the Bloch band separating the two subgaps), and antisymmetric branch
in subgap 2b (it extends into the adjacent Bloch band) may be regarded as
continuations of the, respectively, symmetric branch in subgap 1b, and
antisymmetric one in subgap 2a, which are displayed in Fig. \ref%
{solitons1d_00a}. On the other hand, the origin of the stable branch of
asymmetric solitons in subgap 2b in Fig. \ref{solitons1d_00a} is not clear,
as numerical problems impede to continue it in the direction of decreasing $%
\mu $ and $N$.

\begin{figure}[tbp]
\includegraphics[width=3in]{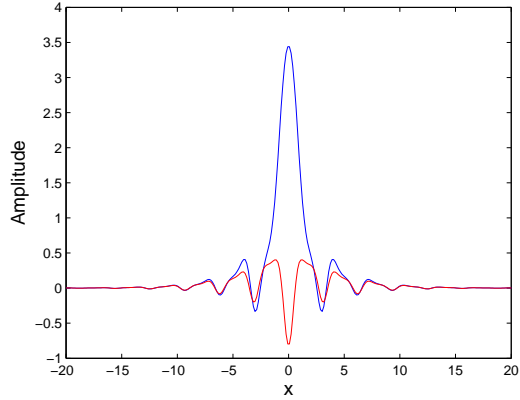}
\caption{(Color online) An asymmetric soliton in the repulsive-repulsive
system, with $\protect\epsilon =8$, $\protect\kappa =1$, $N_{u}=19$, $%
N_{v}=0.9$, $\protect\mu =5.39$.}
\label{solitons_00_profile}
\end{figure}

On the contrary to the situation in the AA model [cf. Figs. \ref%
{solitons1d_11} (a) and (c)], in the present case the symmetric branch does
not undergo any bifurcation, while asymmetric solitons bifurcate
supercritically from the antisymmetric branch. Dependences of asymmetry
ratio $\Theta $ on $\kappa $, and of the bifurcation value, $\kappa _{%
\mathrm{bif}}$, on $N\equiv N_{u}+N_{v}$ for asymmetric solitons in the RR
system are quite similar to those in its AA counterpart, which were shown
above in Fig. \ref{solitons1d_11_properties} (therefore, the dependences for
the RR system are not displayed here). A typical shape of the asymmetric
soliton is presented in Fig. \ref{solitons_00_profile}. It is worthy to note
that both symmetric and antisymmetric soliton branches, but not asymmetric
ones, may become \textit{embedded}, crossing Bloch bands which separate the
subgaps.

\begin{figure}[tbp]
\centering\includegraphics[width=3in]{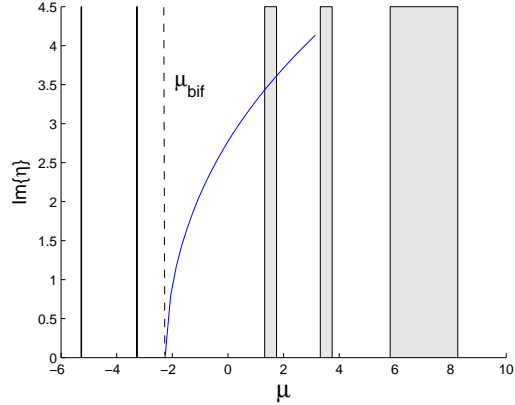}
\caption{(Color online) The instability growth rate for antisymmetric
solitons in the first two (sub)gaps in the repulsion-repulsion model with $%
\protect\kappa =1$ and $\protect\varepsilon =8$. }
\label{solitons1d_00_stability1}
\end{figure}

Similar to what was reported above for the AA model, the reduced eigenvalue
problem based on Eqs. (\ref{symm_reduced_b}) and (\ref{antisymm_reduced_a})
was solved numerically to determine the stability of symmetric and
antisymmetric gap solutions in the RR system (the VK criterion is irrelevant
for self-defocusing models). The analysis demonstrates that the families of
symmetric solitons presented in Figs. \ref{solitons1d_00a} and \ref%
{solitons1d_00b}) are completely stable if the OL is strong enough, but
symmetric solitons may be unstable in a weak lattice (for instance, at $%
\varepsilon =1$). This property is reminiscent of the AA system, where the
antisymmetric solitons are unstable at $\varepsilon =1$, and stable at large
$\varepsilon $, see Figs. \ref{solitons1d_11}, \ref{solitons1d_11_stability}
and \ref{antisymm_stab_thresh}. Further, antisymmetric solitons are stable
before the (anti)symmetry-breaking bifurcation, i.e., at $\mu \leq \mu _{%
\mathrm{bif}}$, and, quite naturally, unstable at $\mu >\mu _{\mathrm{bif}}$
(even if the antisymmetric branch does not seem to be continuously connected
to the bifurcation point, see Fig. \ref{solitons1d_00b}). In the latter
case, the instability growth rate for the antisymmetric solutions is shown
in Fig. \ref{solitons1d_00_stability1}.

Asymmetric solitons bifurcating from the antisymmetric ones are stable
whenever they exist, which was checked by solving the full eigenvalue
problem for them, based on Eqs. (\ref{full_eig_1d}). As well as in the case
of the AA system, these results imply bistability, provided that the OL is
strong enough. Indeed, the symmetric solitons are then stable, and, along
with them, either antisymmetric solitons or asymmetric ones (below or above
the bifurcation, respectively) are stable too.

The predicted stability and instability of symmetric, antisymmetric, and
asymmetric solitons has been verified by direct simulations too. In
particular, the antisymmetric solitons, if unstable, tend to rearrange
themselves into their stable asymmetric counterparts.

\subsection{Twisted solitons and bound states}

Twisted (alias odd) solitons, which feature two out-of-phase amplitude peaks
in one period of the underlying lattice, were found and shown to be stable
in the single-component GPE with attractive nonlinearity \cite{Pelinovsky}
(unlike formally similar \textit{subfundamental} gap solitons in the model
with self-repulsion, which are unstable \cite{Thawatchai}).

\begin{figure}[tbp]
\centering
\subfigure[]{\includegraphics[width=3in]{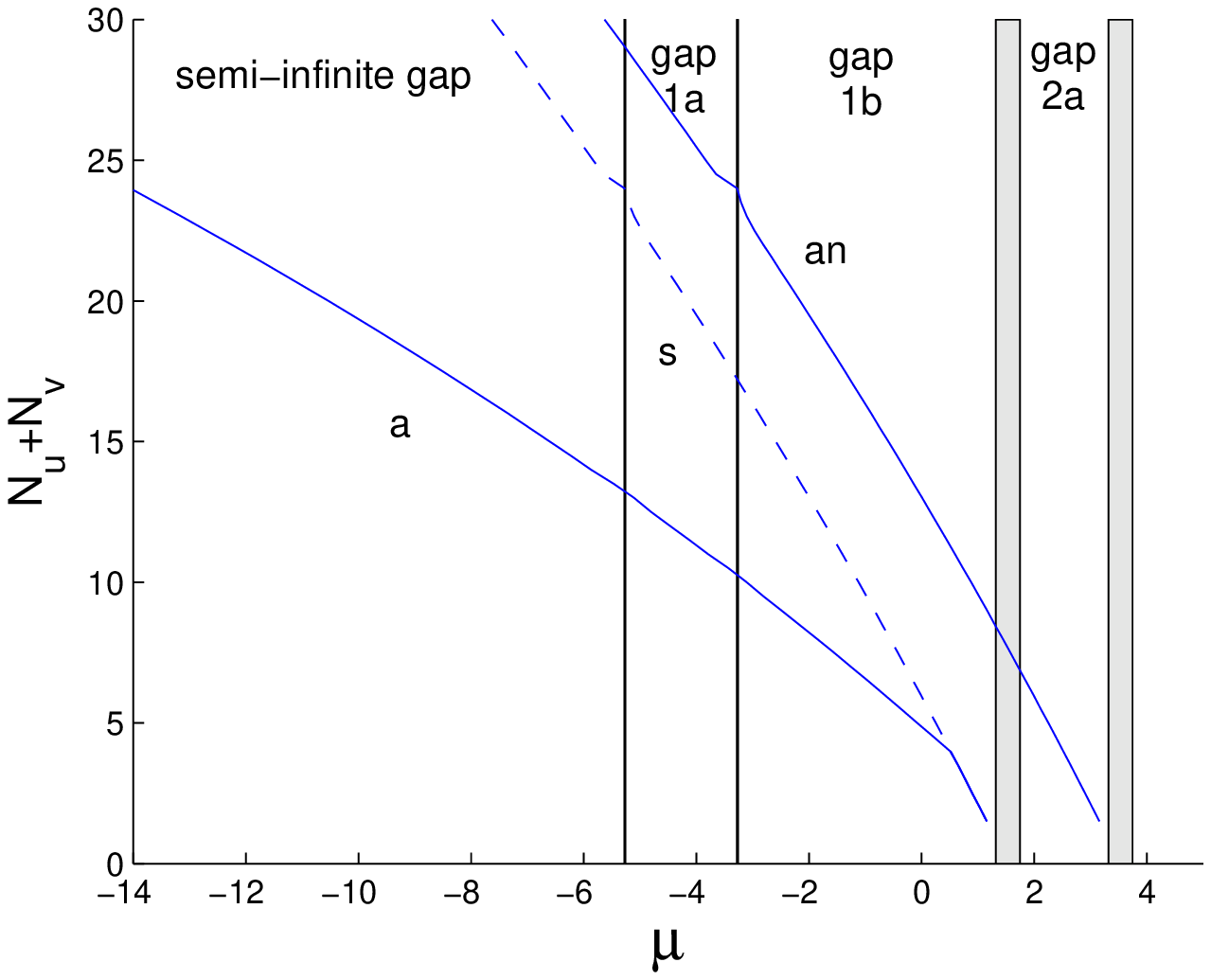}} %
\subfigure[]{\includegraphics[width=3in]{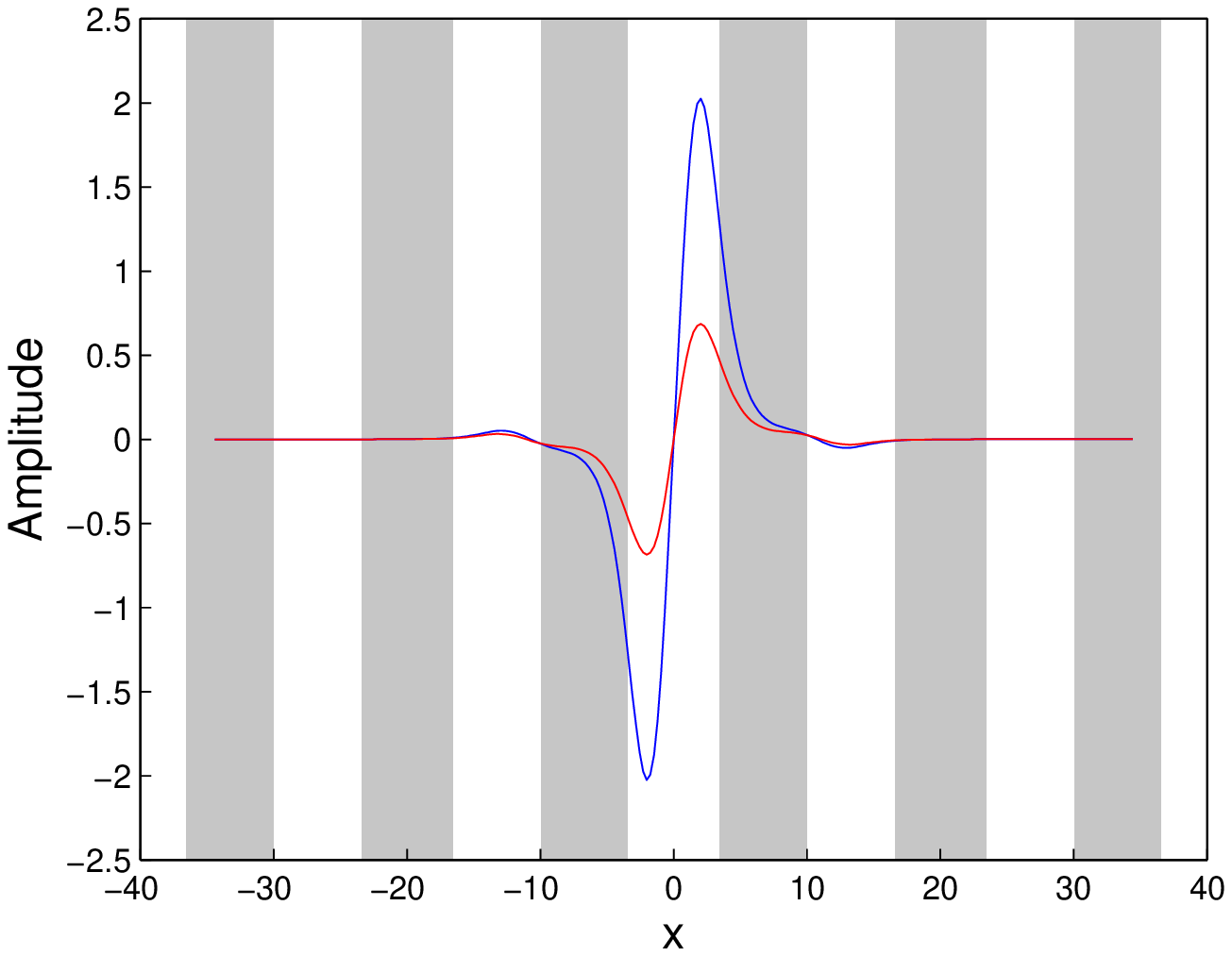}}
\caption{(Color online) (a) Families of twisted (odd) solitons in the
attraction-attraction system, with $\protect\varepsilon =8$ and $\protect%
\kappa =1$. Symmetric twisted solitons are stable up to the bifurcation
point, beyond which they are unstable, while the emerging twisted asymmetric
solitons are stable. Antisymmetric twisted solitons are stable everywhere
(actually, because the lattice is strong enough), cf. Fig. \protect\ref%
{solitons1d_11}(c). (b) An example of an asymmetric twisted soliton with $%
N_{u}=5.3$, $N_{v}=0.7$, $\protect\mu \approx -0.66$. In this figure and in
some figures below, shaded vertical stripes depict the lattice potential
which supports the solitons. }
\label{odd2_solitons}
\end{figure}

We have studied spontaneous symmetry breaking of twisted solitons in the AA
system. Families of twisted solitons in this system are shown in Fig. \ref%
{odd2_solitons}(a). The behavior is generally similar to that of the
fundamental solitons in the same (AA) model: a stable asymmetric branch
bifurcates from the symmetric one, leaving it unstable; however, it is
worthy to note that, in the case shown in Fig. \ref{odd2_solitons}(a), the
bifurcation occurs in the finite (sub)gap, 1b, while the similar bifurcation
of fundamental solitons was observed, in Figs. \ref{solitons1d_11}(a) and
(c), in the semi-infinite gap. The entire antisymmetric branch of the
twisted solitons is stable, provided that the lattice is strong enough, or
unstable otherwise. A typical example of the asymmetric twisted soliton is
shown in Fig. \ref{odd2_solitons}(b).

\begin{figure}[tbp]
\subfigure[]{\includegraphics[width=3in]{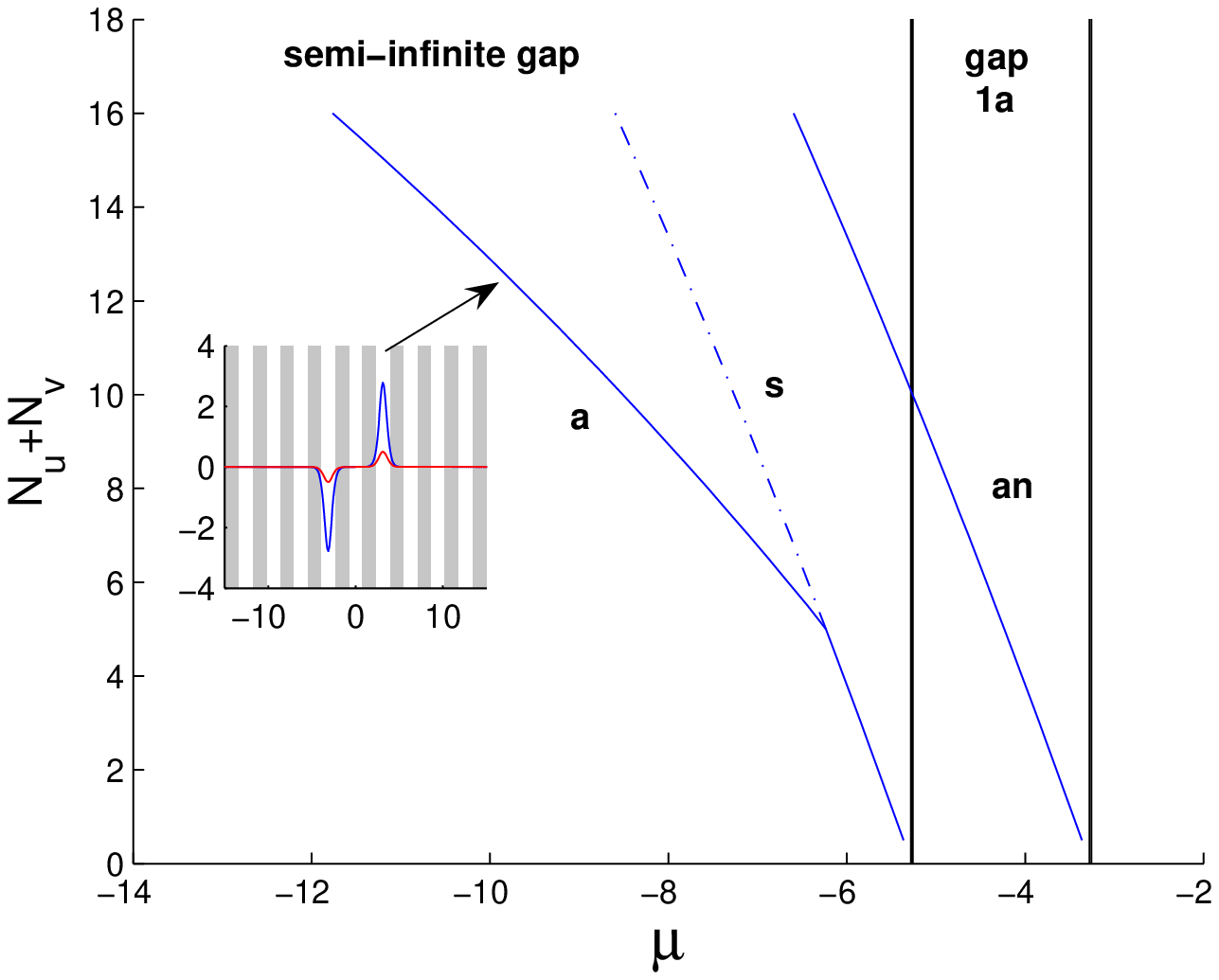}} %
\subfigure[]{\includegraphics[width=3in]{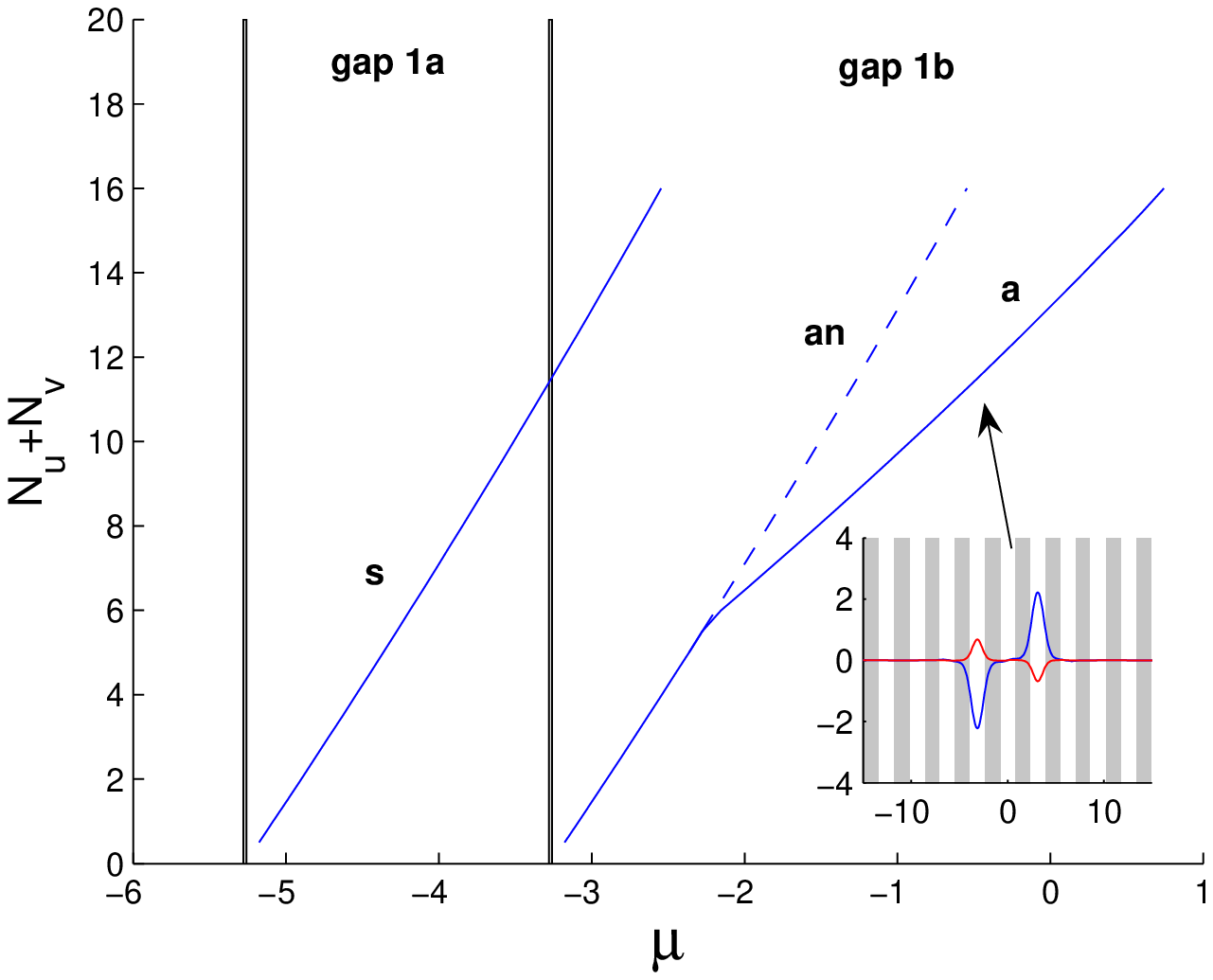}}
\caption{(Color online) Families of twisted bound states, for $\protect%
\varepsilon =8$, $\protect\kappa =1$. Insets show typical bound-soliton
profiles. Panels (a) and (b) pertain to the attraction-attraction and
repulsion-repulsion system, respectively.}
\label{odd_family}
\end{figure}

\begin{figure}[tbp]
\subfigure[]{\includegraphics[width=3in]{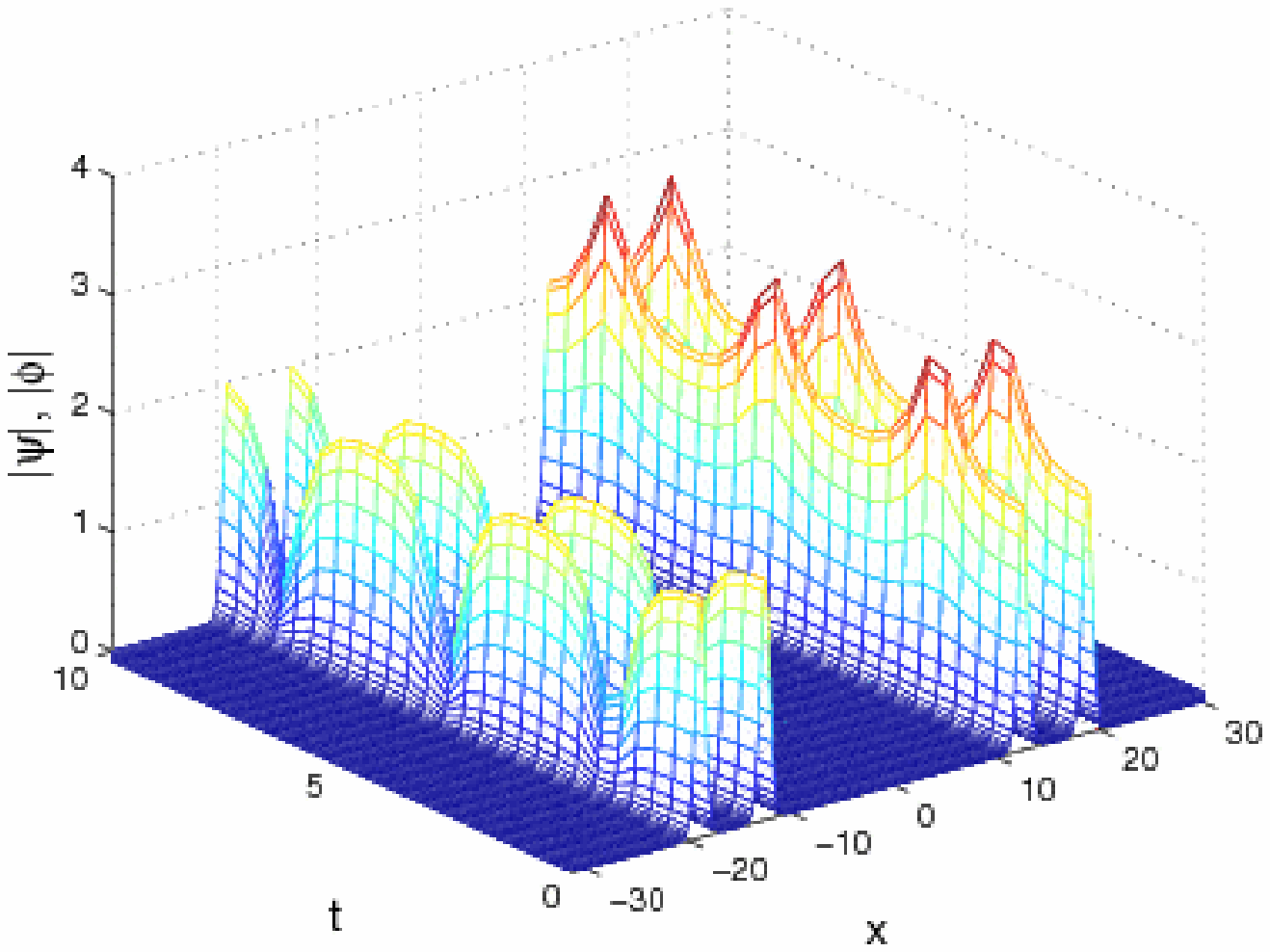}} %
\subfigure[]{\includegraphics[width=3in]{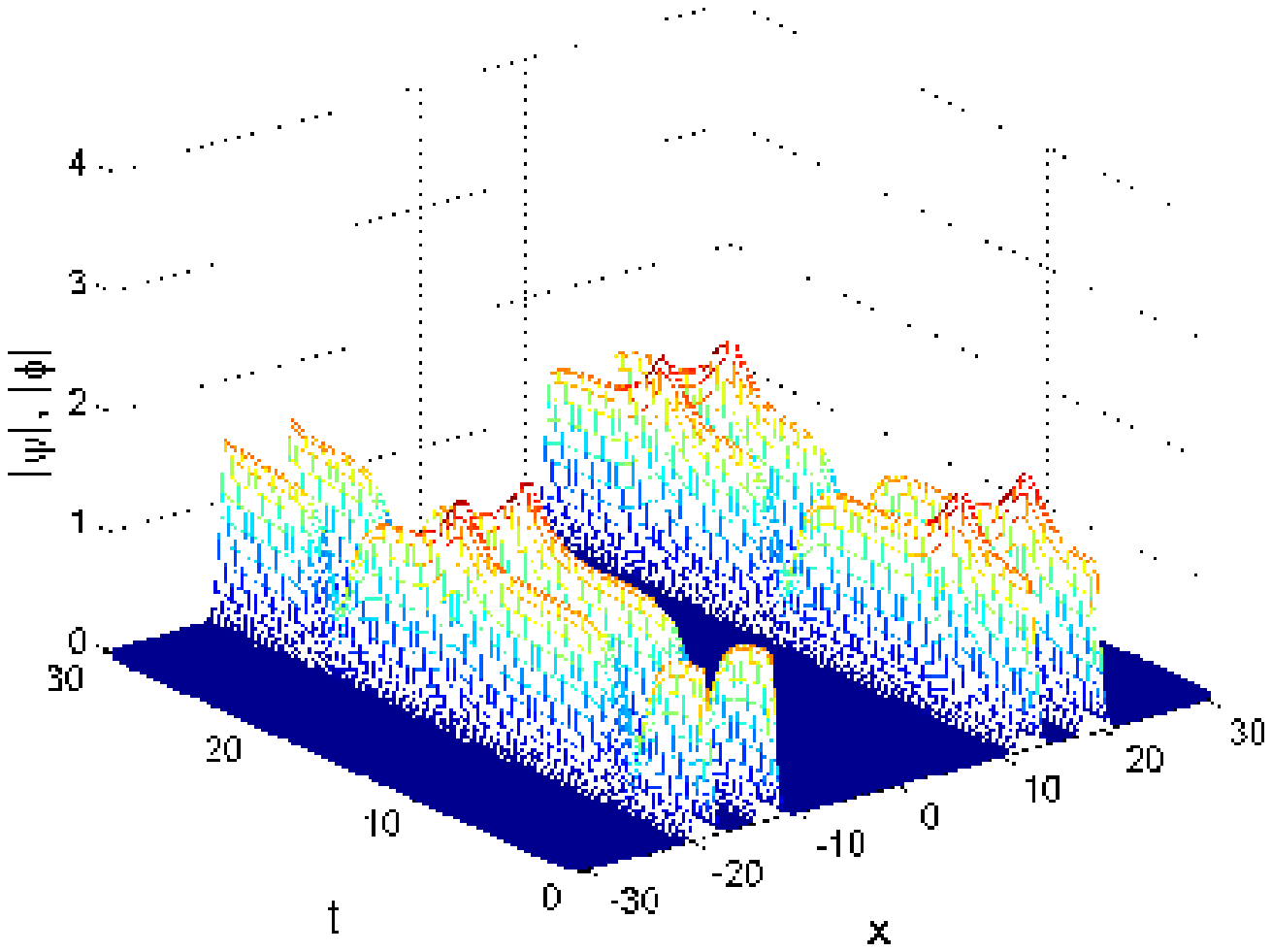}}
\caption{(Color online) Evolution of unstable twisted (odd) bound states of
fundamental solitons in the system with $\protect\varepsilon =8$, $\protect%
\kappa =1$. (a) A bound state of symmetric solitons in the
attraction-attraction system; (b) a bound state of antisymmetric solitons in
the repulsion-repulsion system. In either case, formation of a long-lived
breather is observed.}
\label{odd_evol}
\end{figure}

Bound states of $\pi $-out-of-phase fundamental solitons, which
also feature the odd parity, but include peaks separated by an
empty lattice site, have been found too, in the AA and RR systems
alike (in the single-component GPE with the repulsive
nonlinearity, the presence of an empty site between the peaks is a
necessary condition for the existence of a stable bound state of
two out-of-phase fundamental gap solitons
\cite{Louis,Thawatchai}). Families of the bound states of this
type are shown in Fig. \ref{odd_family}, being quite similar to
fundamental-soliton families, cf. Figs. \ref{solitons1d_11} and
\ref{solitons1d_00a}. A difference is observed in the evolution of
symmetric and anti-symmetric solitons destabilized by the
bifurcation: rather than rearranging into asymmetric solitons,
they give rise to persistent breathers, as shown in Fig.
\ref{odd_evol}.

\section{Asymmetric systems}

As said in Introduction, the two-core system can be made asymmetric by
either assuming that the signs of the scattering lengths are opposite in the
parallel-coupled traps, or by admitting a mismatch between the two OLs. Both
cases are considered in this section.

\subsection{The repulsion-attraction system with aligned lattices}

\begin{figure}[tbp]
\subfigure[]{\includegraphics[width=3in]{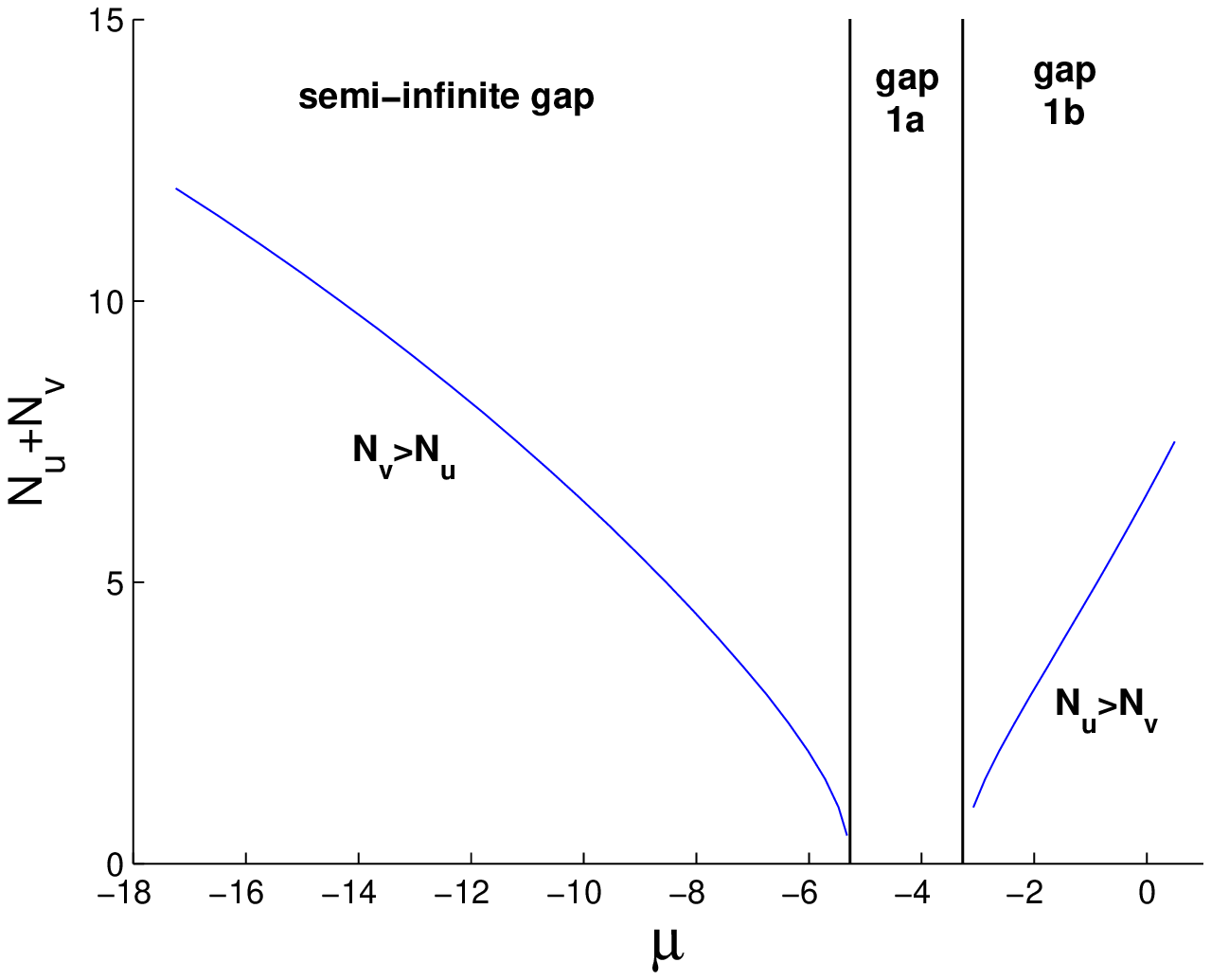}} \subfigure[]{%
\includegraphics[width=3in]{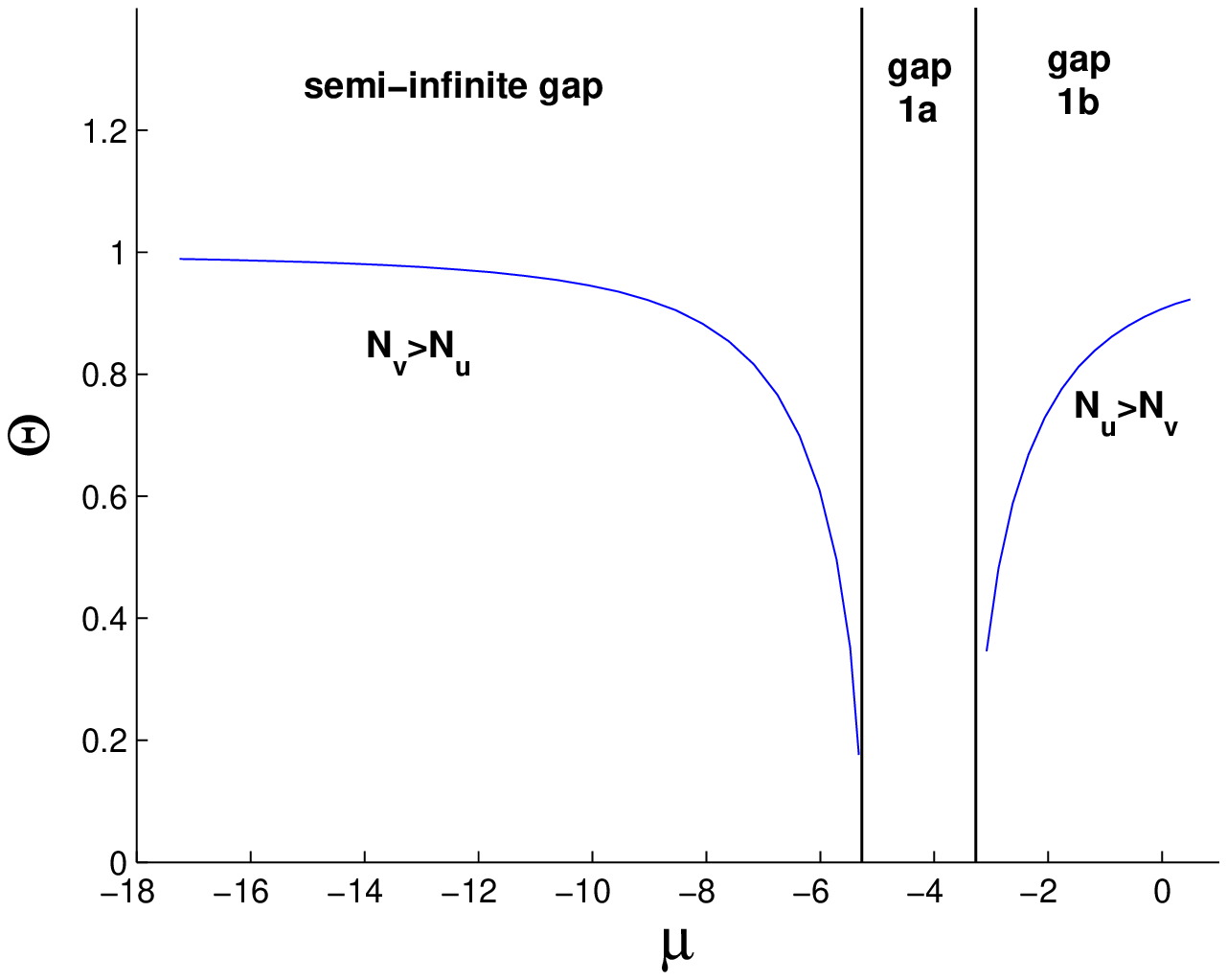}}
\caption{(Color online) Solitons families in the repulsion-attraction model (%
$-g_{1}=g_{2}=1$) with $\protect\varepsilon =8$ and $\protect\kappa =1$.}
\label{solitons1d_10}
\end{figure}

In the RA system (we set $g_{1}=-1$, $g_{2}=1$, assuming opposite signs but
equal absolute values of the nonlinear coefficients in the components),
numerical solution of Eqs. (\ref{uv}) reveals two soliton families, \textit{%
viz}., ones with a dominant repulsive component ($N_{u}>N_{v}$) residing in
subgap 1b, or with a dominant attractive component ($N_{u}<N_{v}$) in the
semi-infinite bandgap (solitons in two-core RA systems, but without the
lattice, were considered in Refs. \cite{Valery}). Generic examples of such
families are presented in Fig. \ref{solitons1d_10}, and typical examples of
the respective soliton shapes are displayed in Fig. \ref%
{solitons1d_10_profile}

\begin{figure}[tbp]
\subfigure[]{\includegraphics[width=3in]{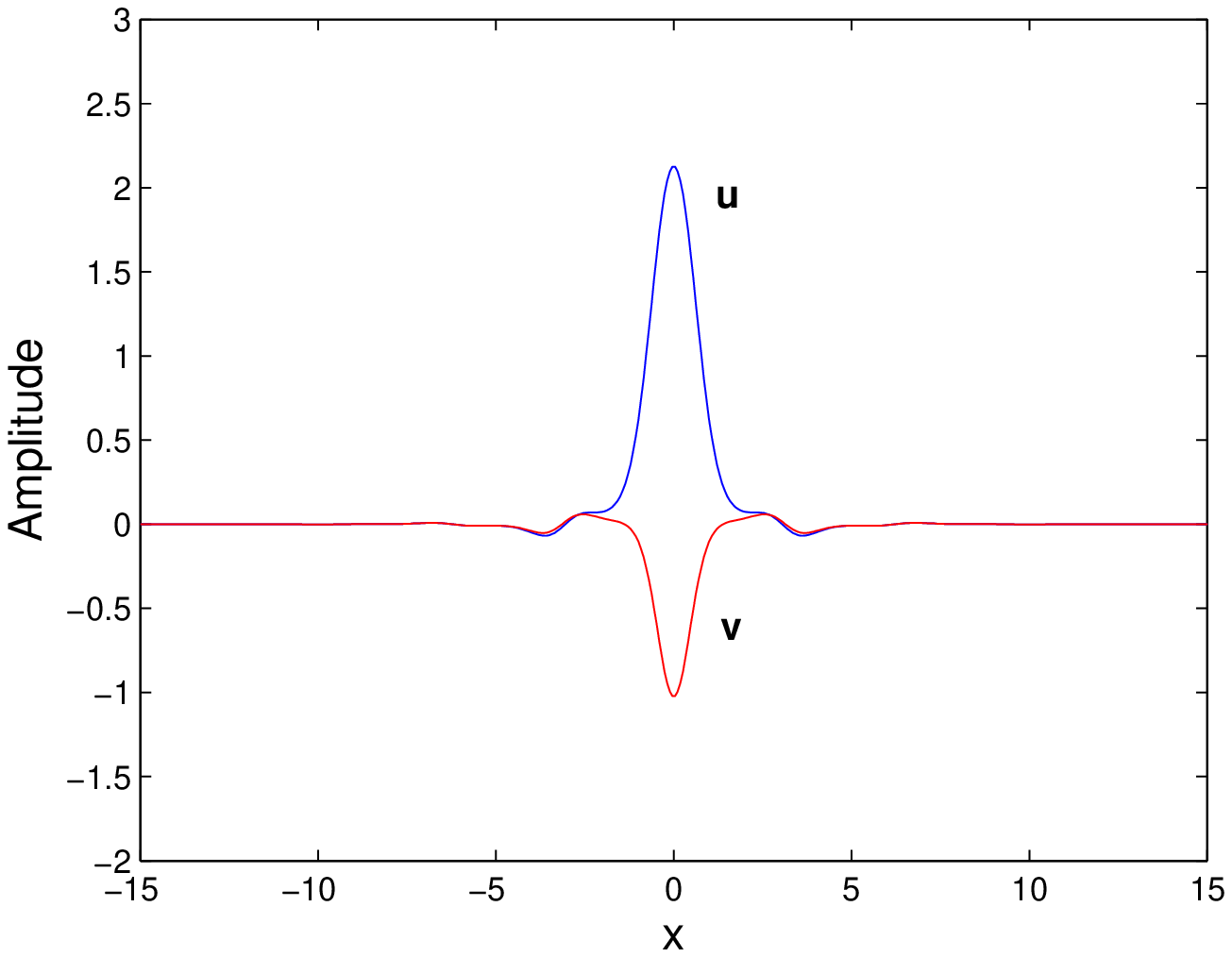}} \subfigure[]{%
\includegraphics[width=3in]{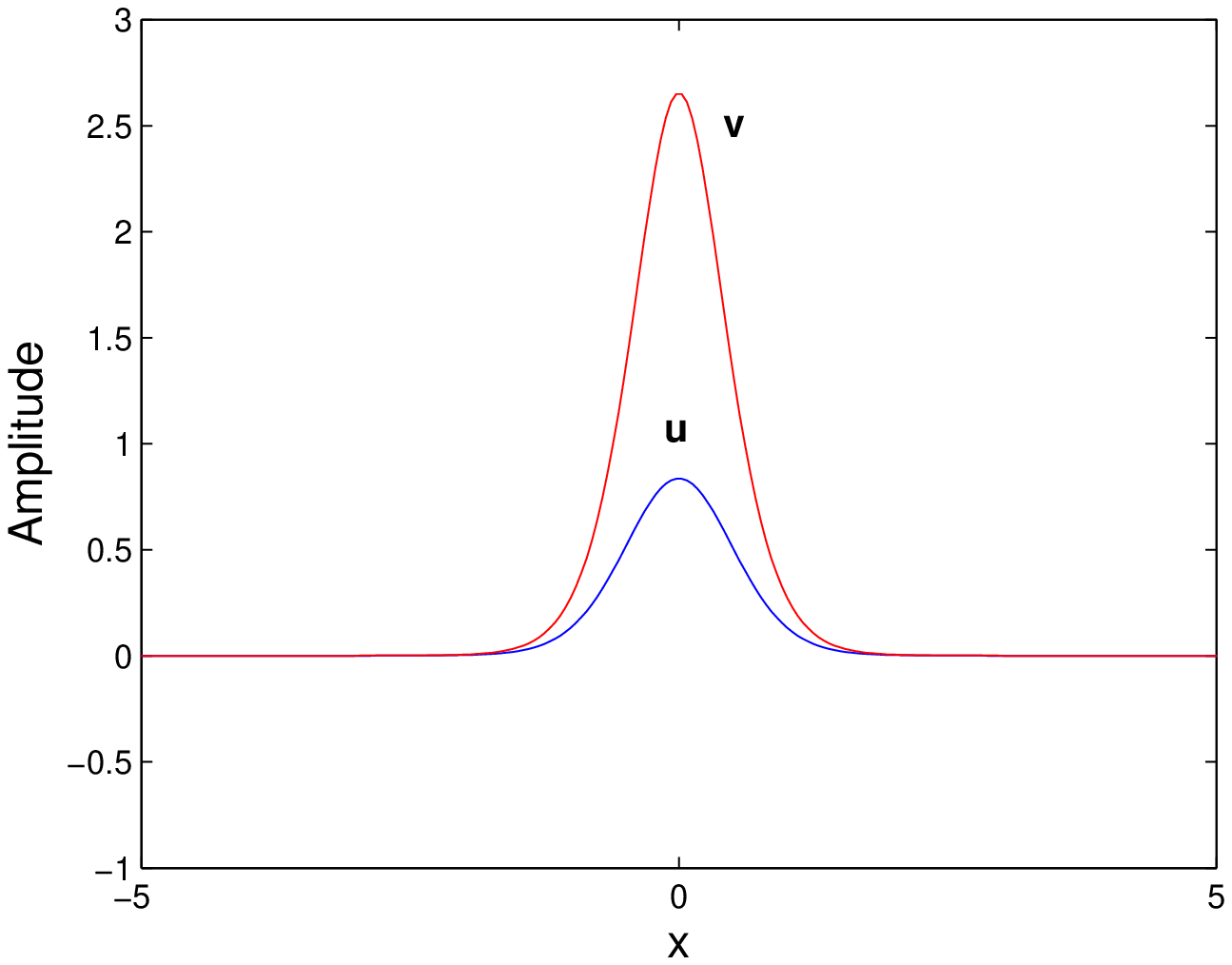}}
\caption{(Color online) Examples of solitons in the repulsion-attraction
model, with $-g_{1}=g_{2}=1$, and $\protect\varepsilon =8$, $\protect\kappa %
=2$. (a) $N_{u}=5.1$, $N_{v}=0.9$, $\protect\mu \approx -0.105$; (b) $%
N_{u}=0.6$, $N_{v}=5.4$, $\protect\mu \approx -9.58$. In either case, the
total norm is $N=6$.}
\label{solitons1d_10_profile}
\end{figure}

The computation of the eigenvalues for perturbation modes demonstrates that
both families are \emph{completely stable}. Note that the stability of the
attraction-dominated solutions, with $N_{u}<N_{v}$, is also suggested by the
VK criterion, if applied to dependence $N(\mu )$ in Fig. \ref{solitons1d_10}%
(a) (for the repulsion-dominated solutions, with $N_{u}>N_{v}$, the VK
criterion is irrelevant, as mentioned above).

\subsection{Bifurcations and \textit{pseudo-bifurcations} in systems with
mismatched lattices}

To consider effects of the mismatch between the OLs in the two cores, we
will here focus on Eqs. (\ref{model_1d}) with $\Delta =\pi /2$ and $\Delta
=\pi $ (the latter value corresponding to the largest mismatch, similar to a
system of parallel Bragg gratings with the maximum mismatch, that was
recently considered in Ref. \cite{Sukhorukov}). For all types of the
nonlinear interactions in the cores, at $\kappa =0$ the solitons in the
decoupled shifted lattices were taken as $u(x)=\hat{u}_{0}(x),v(x)=0$, where
$\hat{u}_{0}(x)$ is the corresponding soliton in the single-component GPE.
At this point, the asymmetry ratio defined in Eq. (\ref{theta}) is at its
maximum, $\Theta =1$. As the linear-coupling coefficient, $\kappa $,
increases, the solution becomes less asymmetric.

\begin{figure}[tbp]
\subfigure[]{\includegraphics[width=3in]{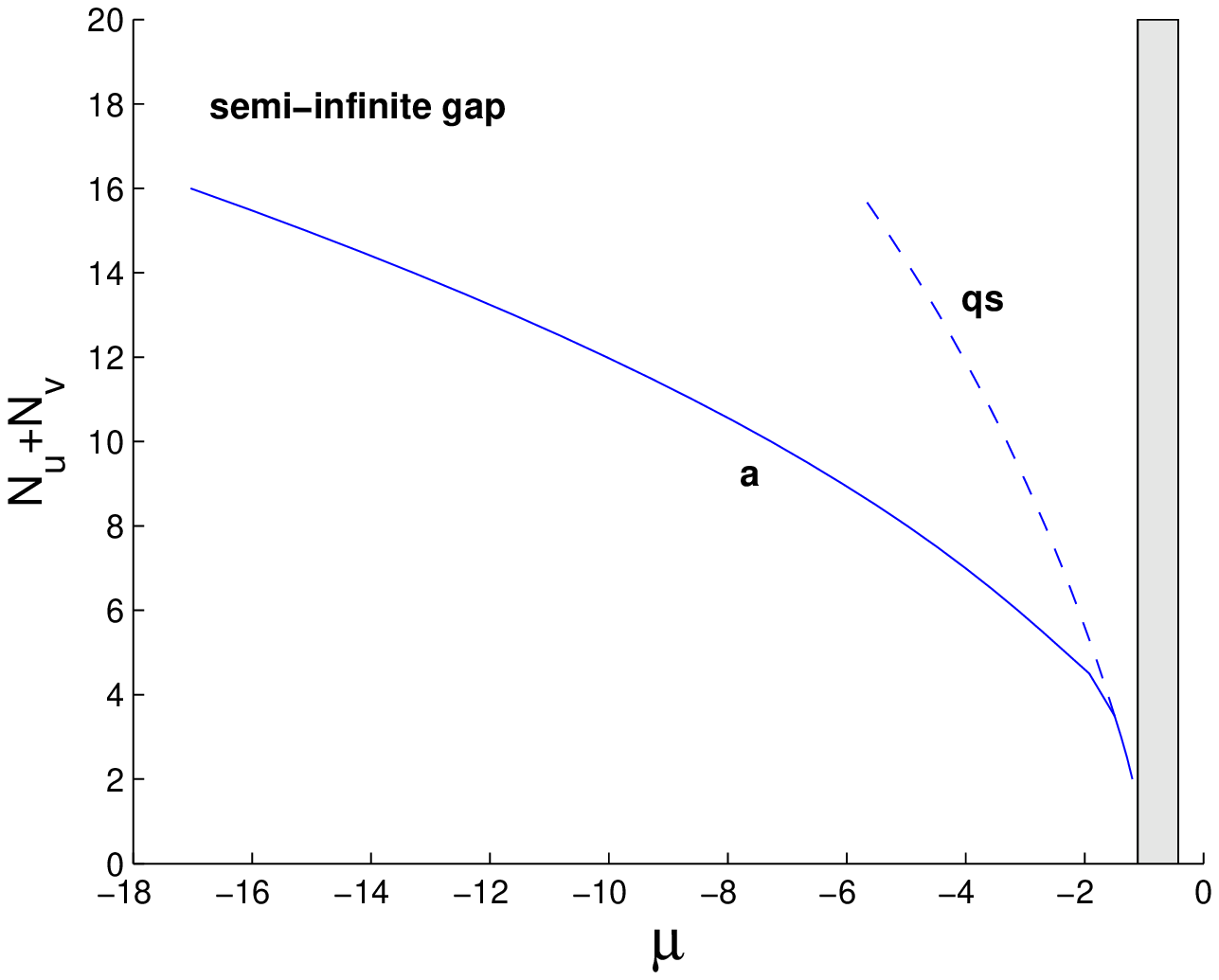}}
\subfigure[]{\includegraphics[width=3in]{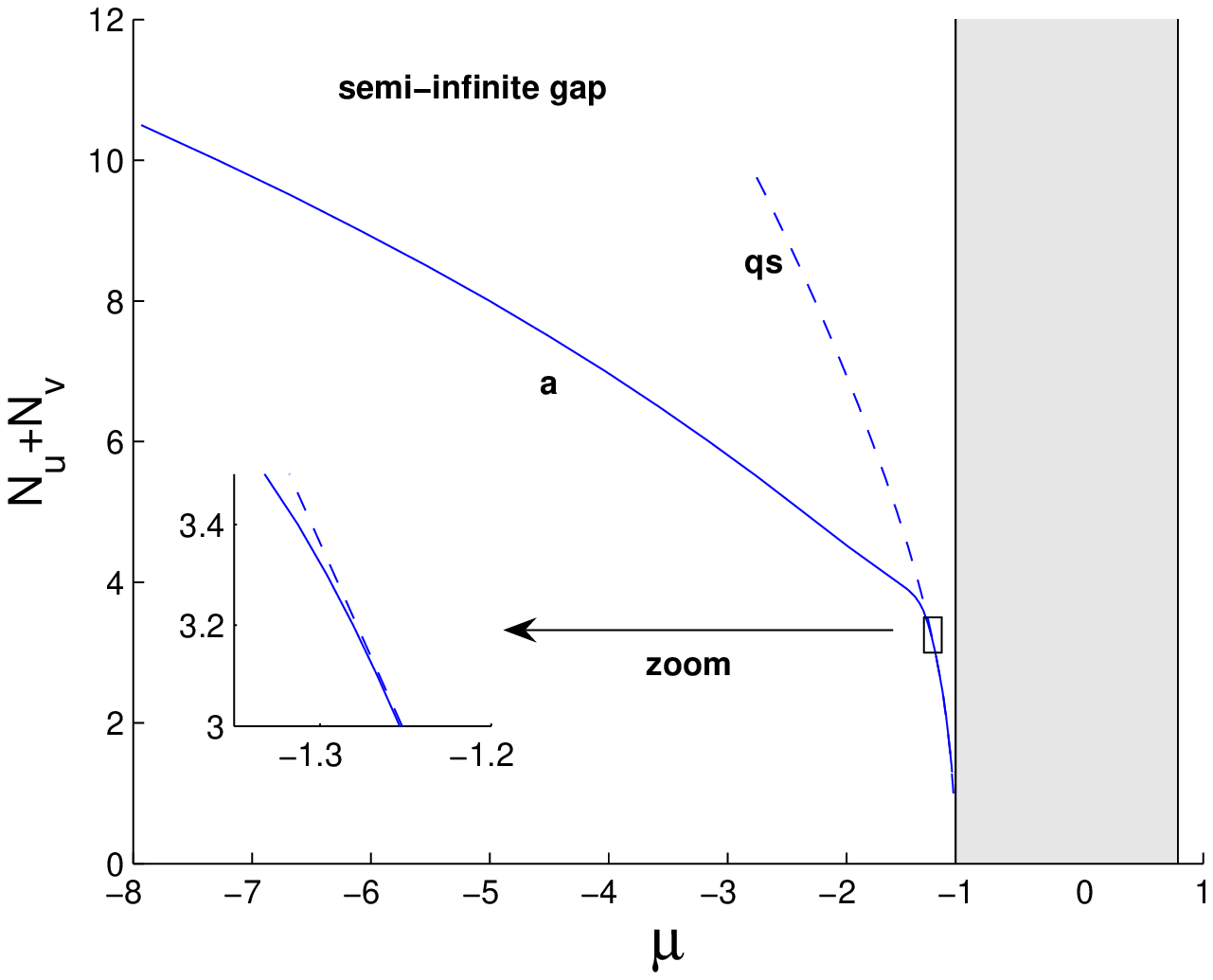}
\label{shifted11_family_b}}
\caption{(Color online) Solitons families in the attraction-attraction
system with mismatched lattices for $\protect\varepsilon =1$ and $\Delta =%
\protect\pi /2$ (a) and $\Delta =\protect\pi $ (b). Labels ``a"
and ``qs" refer to asymmetric and quasi-symmetric branches,
respectively. } \label{shifted11_family}
\end{figure}

Figure \ref{shifted11_family} displays soliton families found in the AA
system. As in Ref. \cite{Yossi}, which treated a mismatched system of
parallel-coupled Bragg gratings, we distinguish between asymmetric and
quasi-symmetric (QS) solitons [a branch of quasi-antisymmetric (QAS)
solutions was found too but is not shown, as it is completely unstable, cf.
Fig. \ref{solitons1d_11}]. QS solitons have equal norms in both components,
i.e., $\Theta =0$, and feature similar, although not identical, profiles of
the components. The bifurcation chart for the AA system with $\Delta =\pi /2$
is very similar to the one for the same system with aligned OLs ($\Delta =0$%
), the symmetric solutions being replaced by their QS counterparts. In
particular, the asymmetric branch bifurcates from the QS one.

\begin{figure}[tbp]
\subfigure[]{\includegraphics[width=3in]{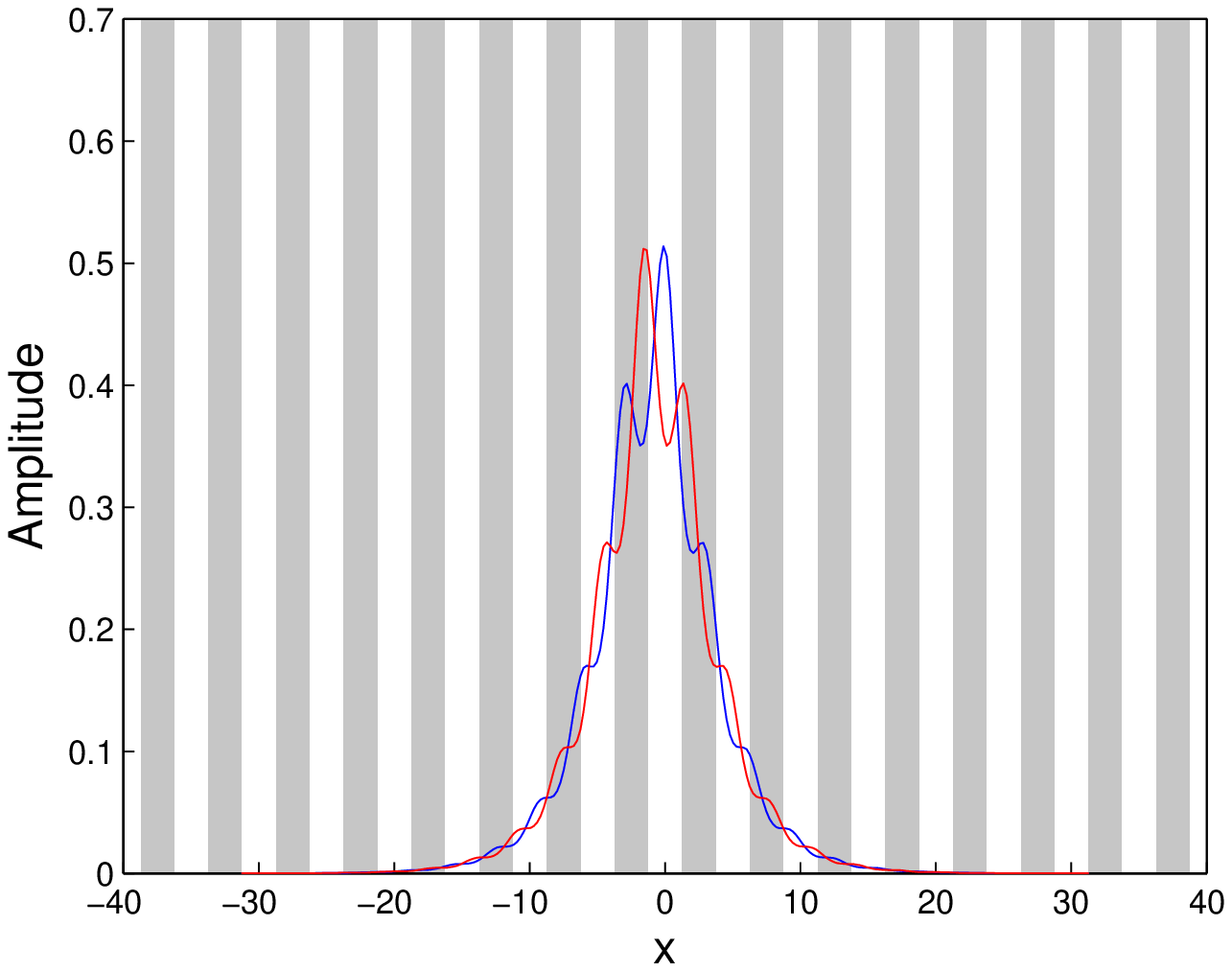}} %
\subfigure[]{\includegraphics[width=3in]{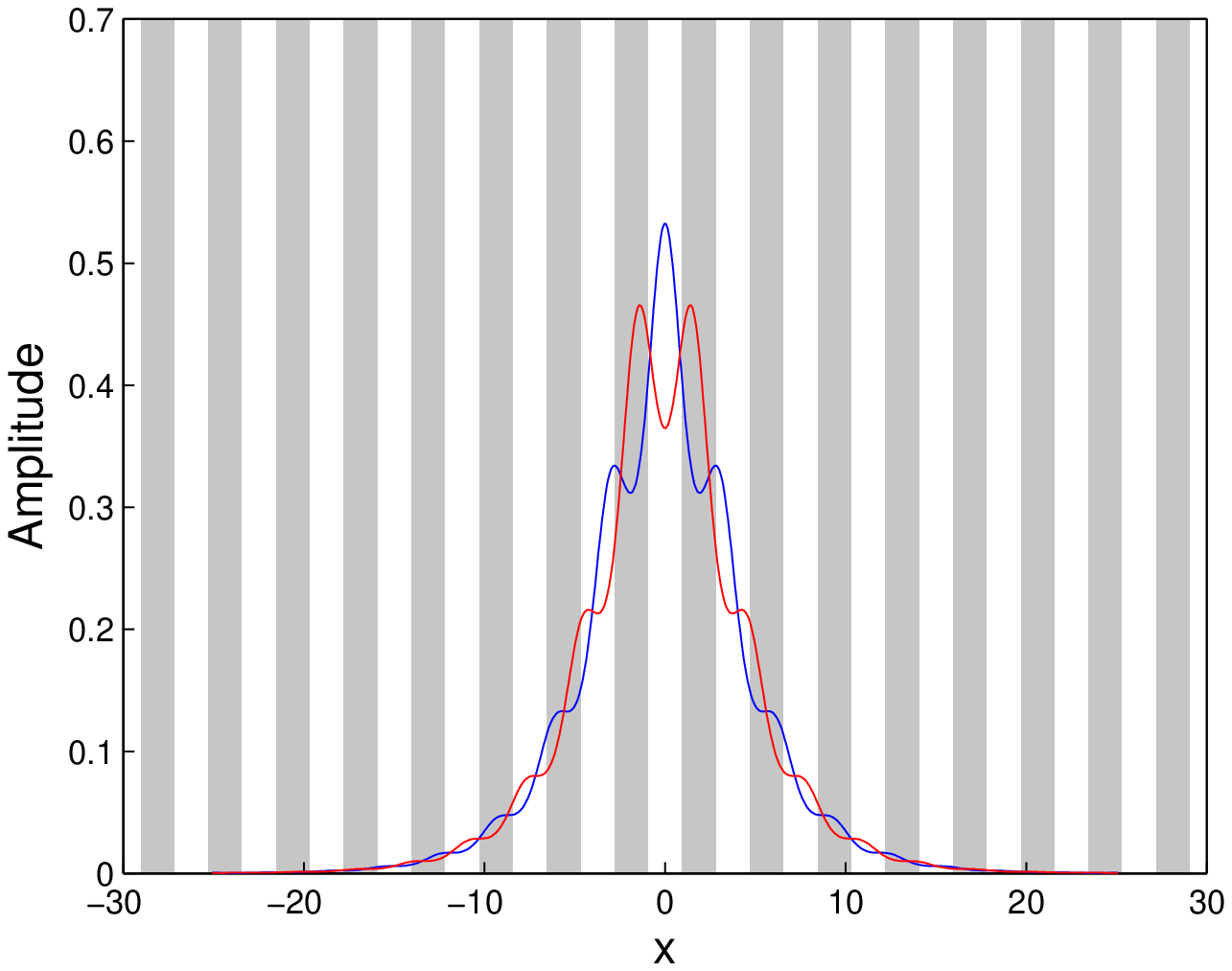}}
\caption{(Color online) Comparison of soliton profiles in the
attraction-attraction model with $\Delta =\protect\pi $ and $\protect%
\varepsilon =1$, $\protect\kappa =1$. (a) A quasi-symmetric soliton, with $%
\Theta =0$; (b) a slightly asymmetric soliton, with $\Theta \approx 10^{-4}$%
. Both solitons have equal norms, $N=3.5$. Here and below, the set of
vertical shaded stripes represents the periodic lattice potential in the
first core.}
\label{no_bif_profiles}
\end{figure}

At first glance, it may seem that, in the AA system with $\Delta =\pi $, the
branch of asymmetric solitons also bifurcates from the QS one. However, this
is not the case. A blow-up (inset in Fig. \ref{shifted11_family}b) shows a
\textit{pseudo-bifurcation}, which means that asymmetric and QS branches
gradually approach each other to a point where they seem indistinguishable,
but they never merge, therefore the QS solutions always remain (strictly
speaking) unstable, while the asymmetric ones are always stable (the
stability is discussed in more detail below). In Fig \ref{no_bif_profiles},
we display an example of comparison between shapes of asymmetric and QS
solitons with equal norms, $N=3.5$, when they are very close to each other,
the asymmetric one having $\Theta \approx 10^{-4}$ (its QS counterpart has $%
\Theta $ exactly equal to zero). Note that the centers of the two components
in the QS soliton are slightly separated, while in the asymmetric soliton
they exactly coincide.

\begin{figure}[tbp]
\subfigure[]{\includegraphics[width=3in]{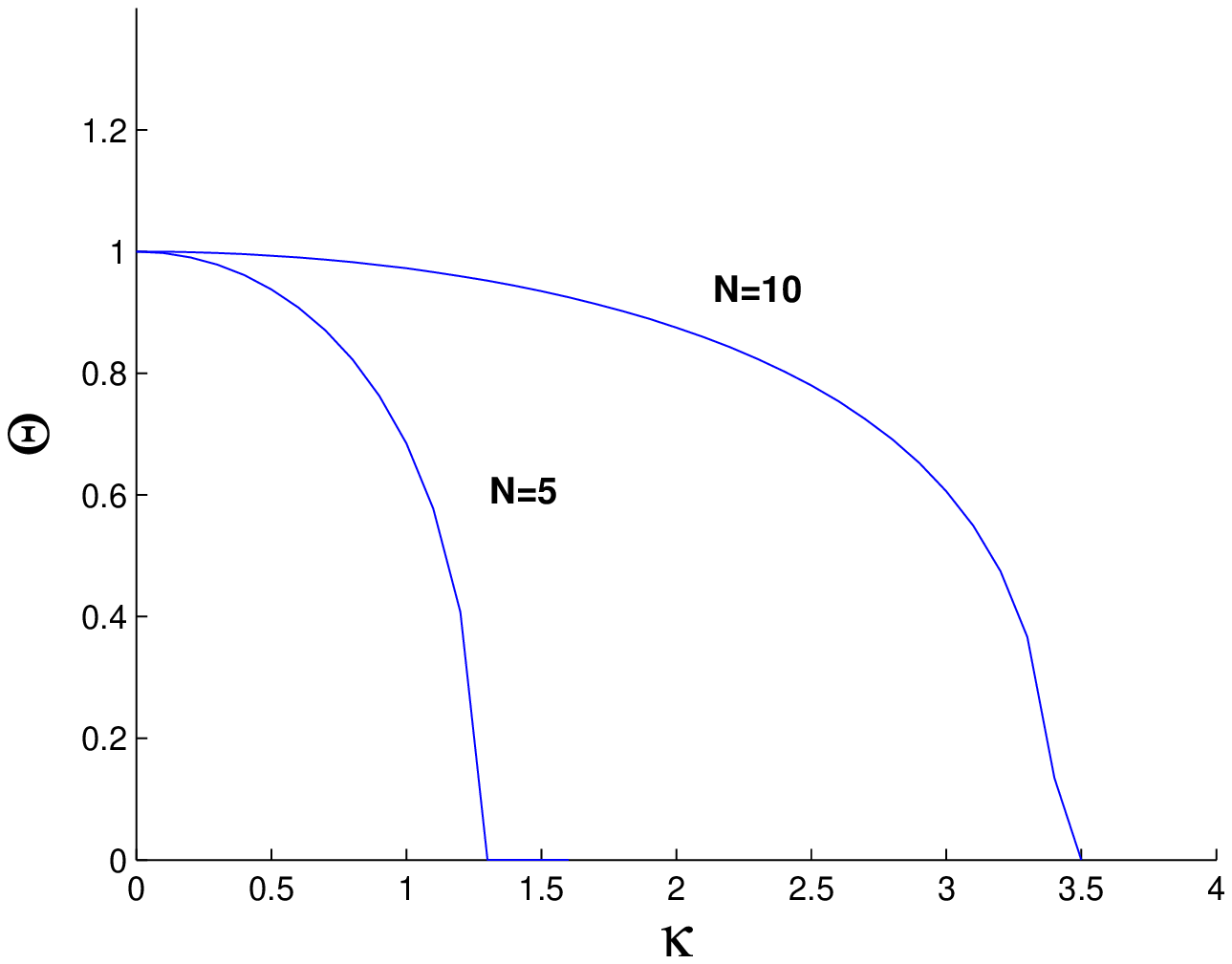}} %
\subfigure[]{\includegraphics[width=3in]{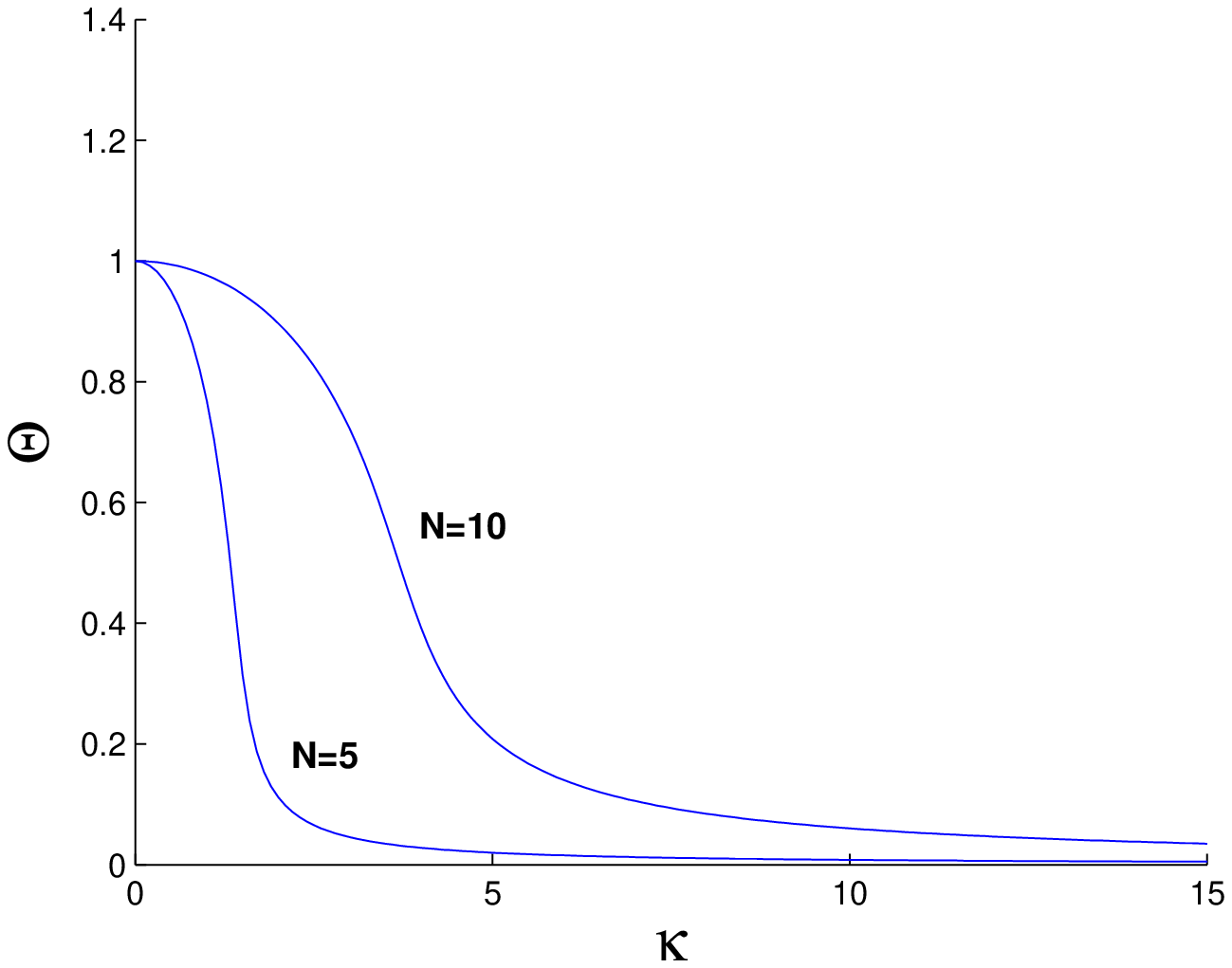}}
\caption{(Color online) The symmetry ratio for solitons in the mismatched
attraction-attraction model with $\protect\varepsilon =1$ versus the
coupling coefficient, $\protect\kappa $: (a) $\Delta =\protect\pi /2$; (b) $%
\Delta =\protect\pi $ (similar dependences for the aligned system, with $%
\Delta =0$, are displayed in Fig. \protect\ref{solitons1d_11_properties}).}
\label{shifted11_1}
\end{figure}

Figure \ref{shifted11_1} additionally illustrates the difference between the
true quasi-symmetry-breaking bifurcation in the AA system with $\Delta =\pi
/2$, and the pseudo-bifurcation in the system with $\Delta =\pi $. In the
latter case, the branches closely approach but never merge, cf. Fig. \ref%
{shifted11_family}.

\begin{figure}[tbp]
\subfigure[]{\includegraphics[width=3in]{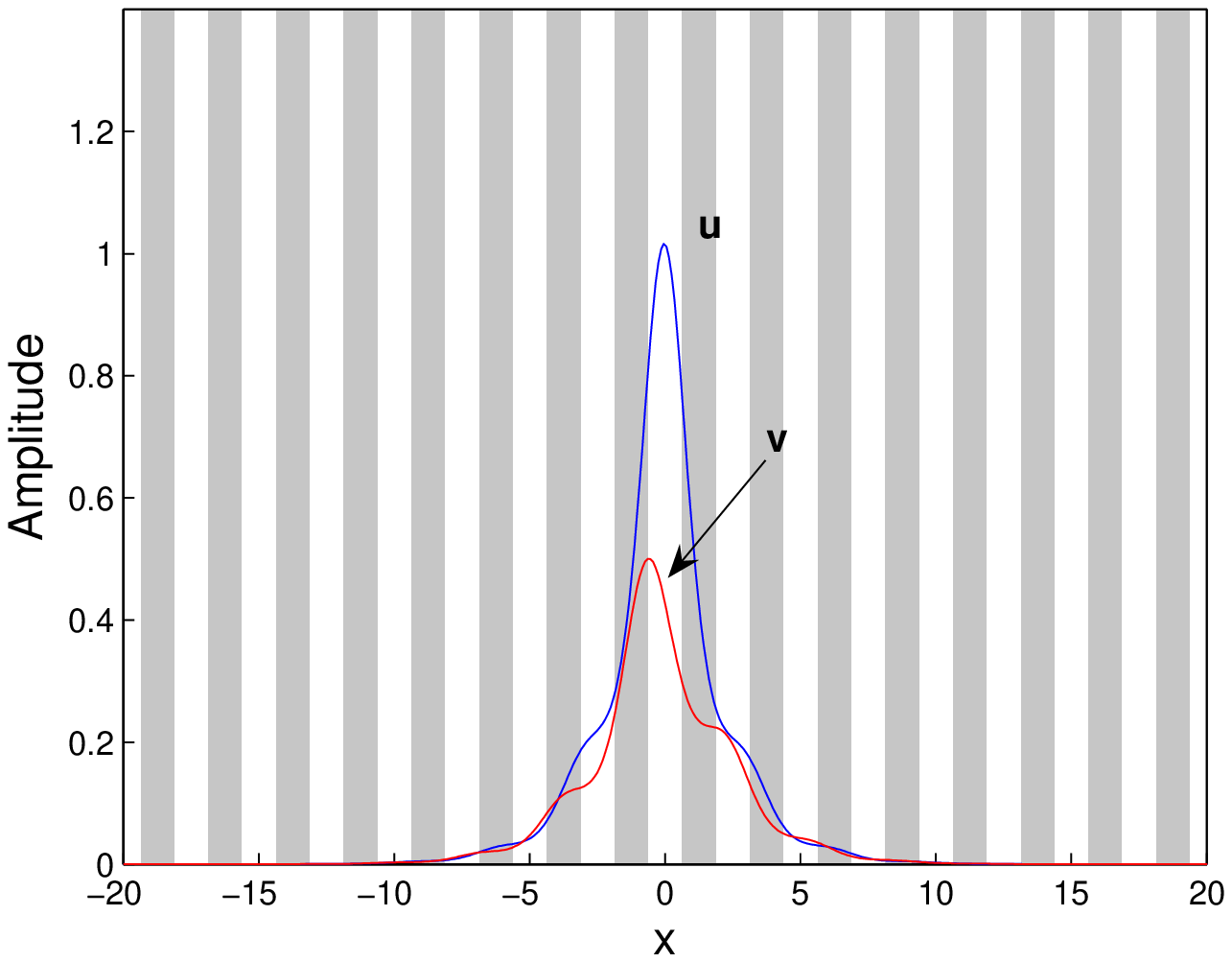}}
\subfigure[]{\includegraphics[width=3in]{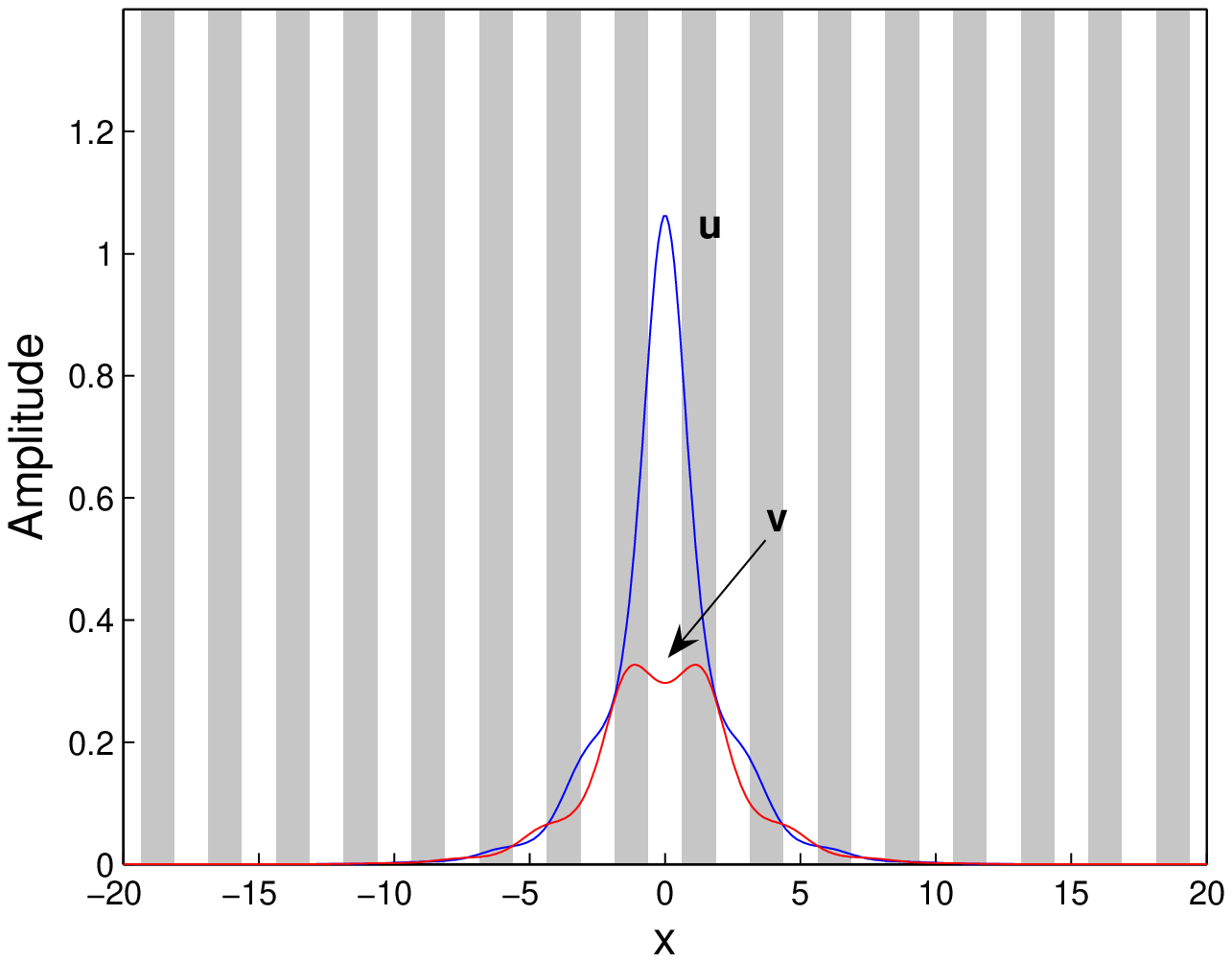}}
\caption{(Color online) Typical asymmetric solitons in the mismatched
attraction-attraction system with $\protect\varepsilon =1$: (a) $\Delta =%
\protect\pi /2$, $N=2.5$, $\Theta =0.51$, $\protect\mu =-0.83$, $\protect%
\kappa =0.4$; (b) $\Delta =\protect\pi $, $N=2.5$, $\Theta=0.63$, $\protect%
\mu =-0.85$, $\protect\kappa =0.4$.}
\label{shifted11_2}
\end{figure}

Typical shapes of strongly asymmetric solitons in the mismatched AA model
are displayed in Fig. \ref{shifted11_2}.
Naturally, the solitons are more asymmetric at $\Delta =\pi $ .

\begin{figure}[tbp]
\subfigure[]{\includegraphics[width=3in]{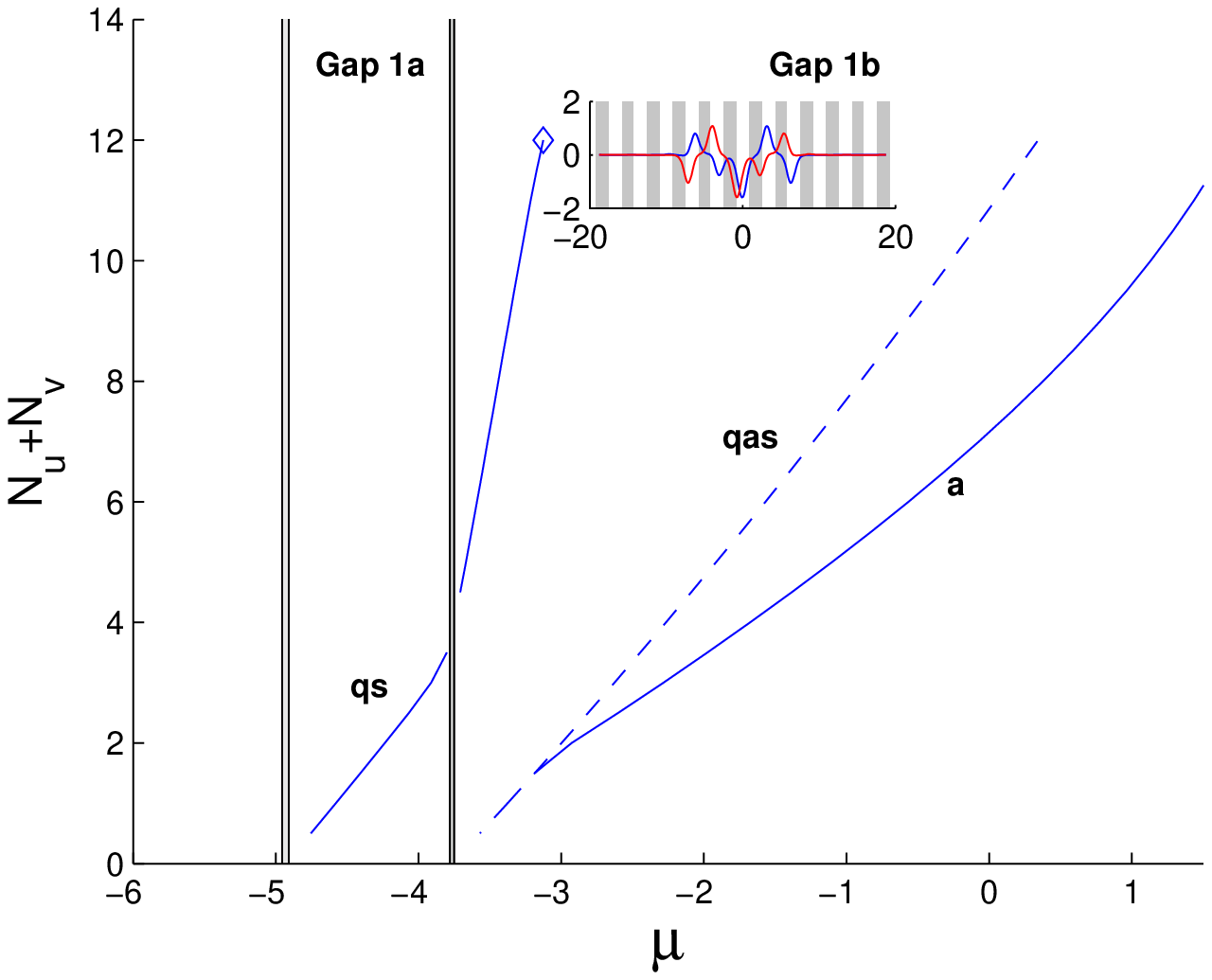}} %
\subfigure[]{\includegraphics[width=3in]{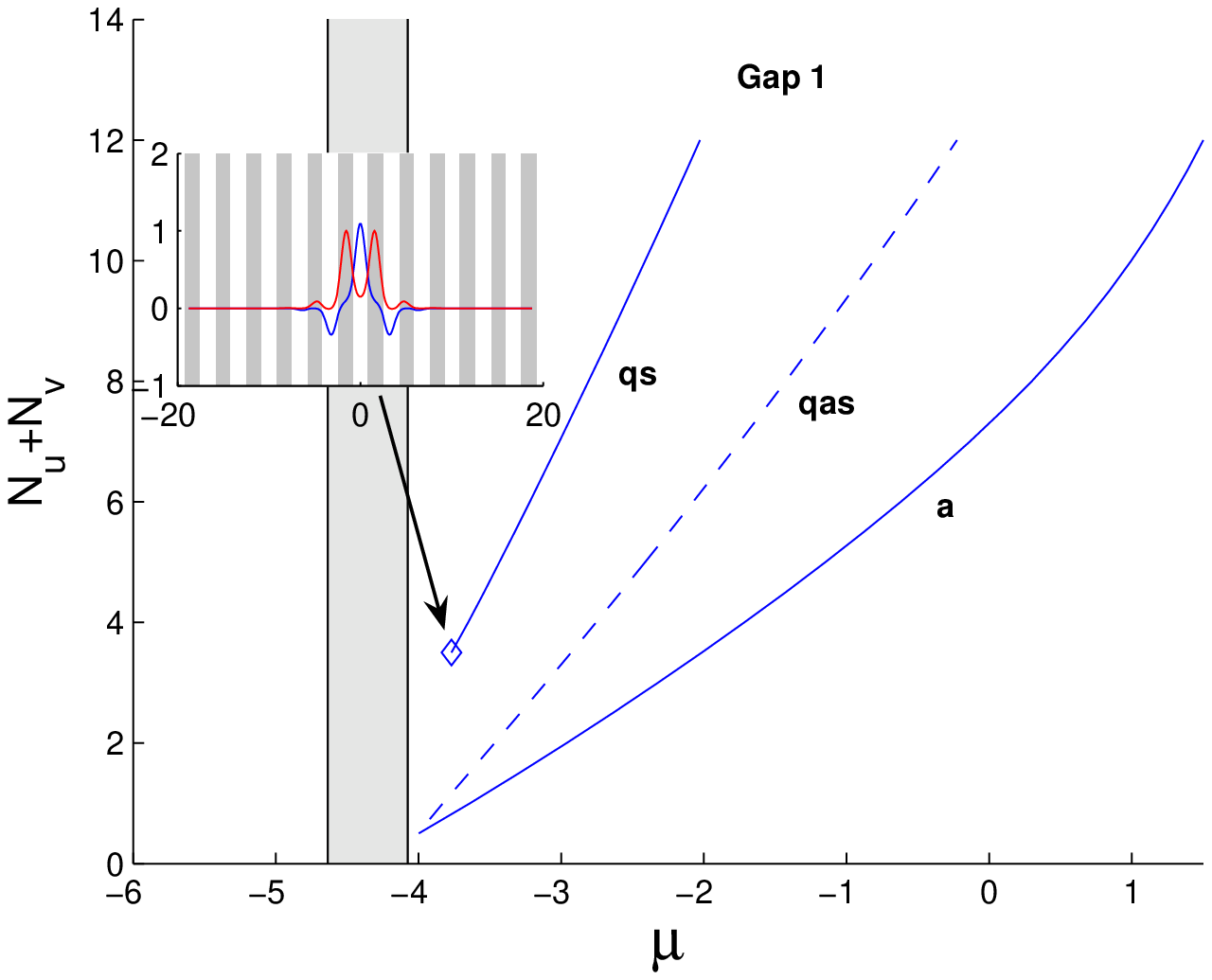}}
\caption{(Color online) Soliton families in the misaligned
repulsion-repulsion system, with $\protect\varepsilon =8$. Labels
``a", ``qs", and ``qas" refer to asymmetric, quasi-symmetric and
quasi-antisymmetric branches, respectively. Insets show soliton
profiles at points marked by diamonds. (a) $\Delta =\protect\pi
/2$; (b) $\Delta =\protect\pi $.} \label{shifted00_1}
\end{figure}

Families of solitons found in the mismatched RR system are presented in Fig. %
\ref{shifted00_1}. The QS branch is called this way only in the sense that
peak amplitudes of the two components have the same sign. In fact, profiles
of the components in the QS soliton may be very different, as insets show in
Fig. \ref{shifted00_1}. The quasi-antisymmetry-breaking bifurcation in the
RR system and respective pseudo-bifurcation are illustrated by Fig. \ref%
{shifted00_3}, and a set of typical profiles of the solitons is displayed in
Fig. \ref{shifted00_2}.

\begin{figure}[tbp]
\subfigure[]{\includegraphics[width=3in]{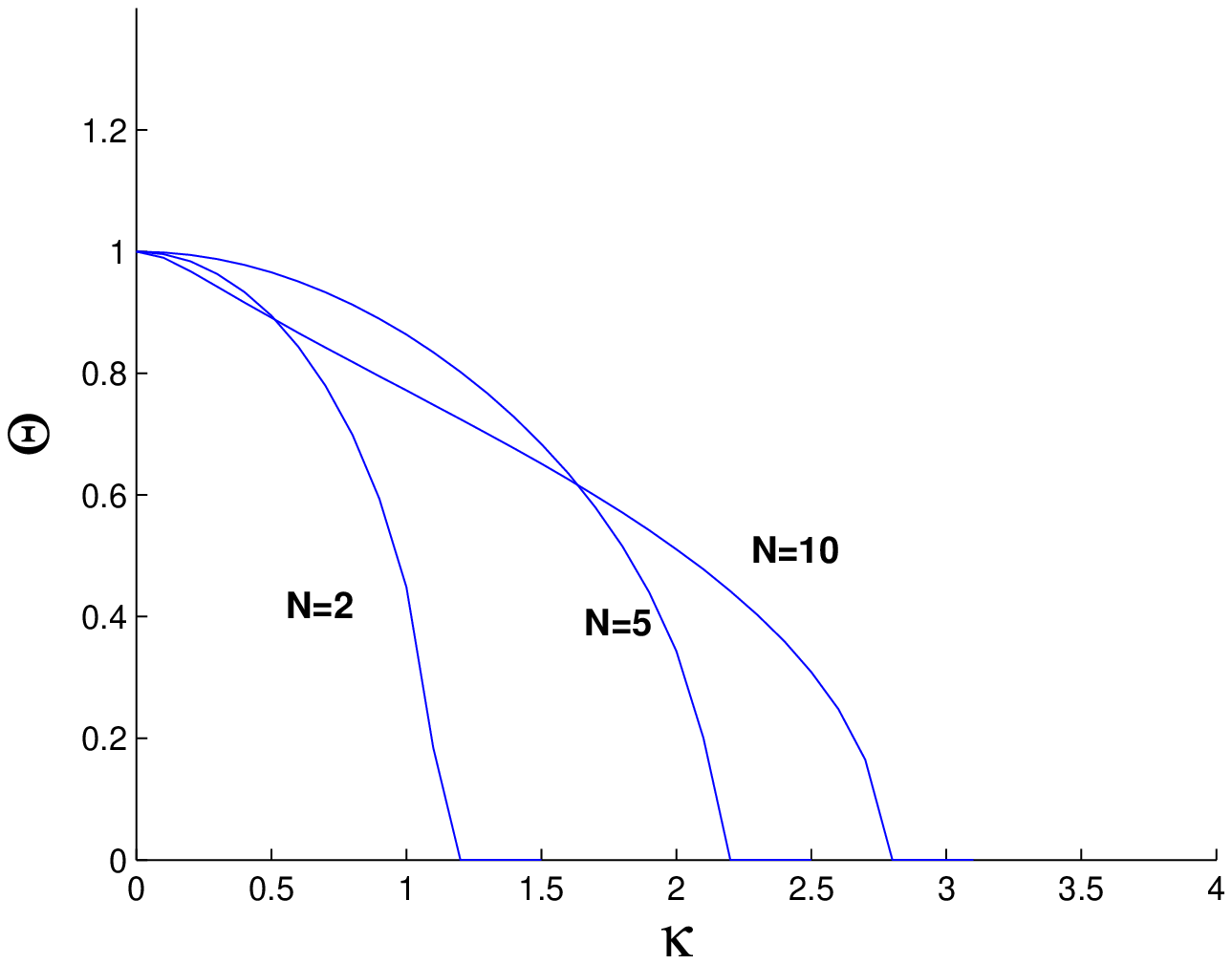}} %
\subfigure[]{\includegraphics[width=3in]{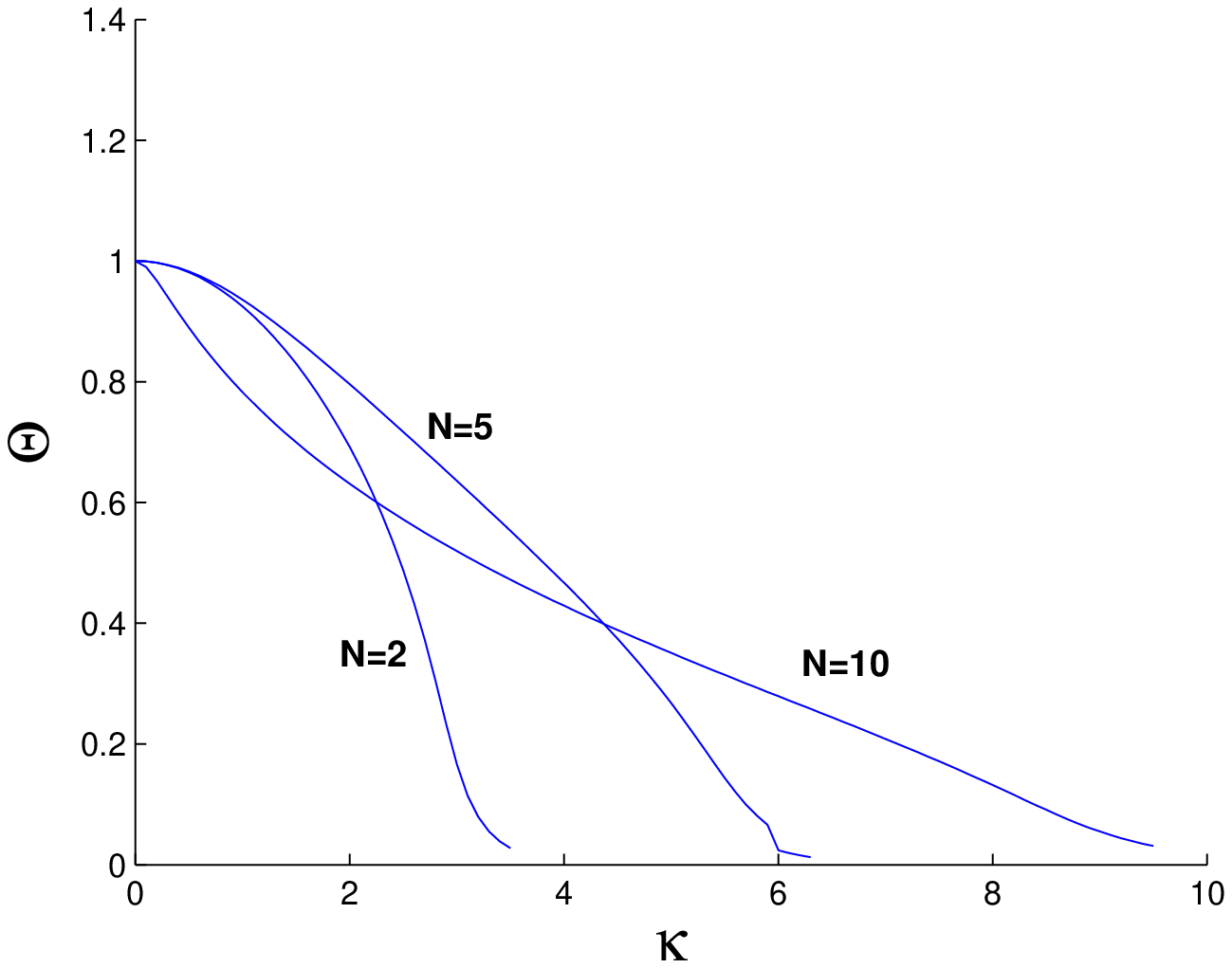}}
\caption{(Color online) The asymmetry ratio of solitons in the mismatched
repulsion-repulsion system with $\protect\varepsilon =8$: (a) $\Delta =%
\protect\pi /2$; (b) $\Delta =\protect\pi $.}
\label{shifted00_3}
\end{figure}

\begin{figure}[tbp]
\subfigure[]{\includegraphics[width=3in]{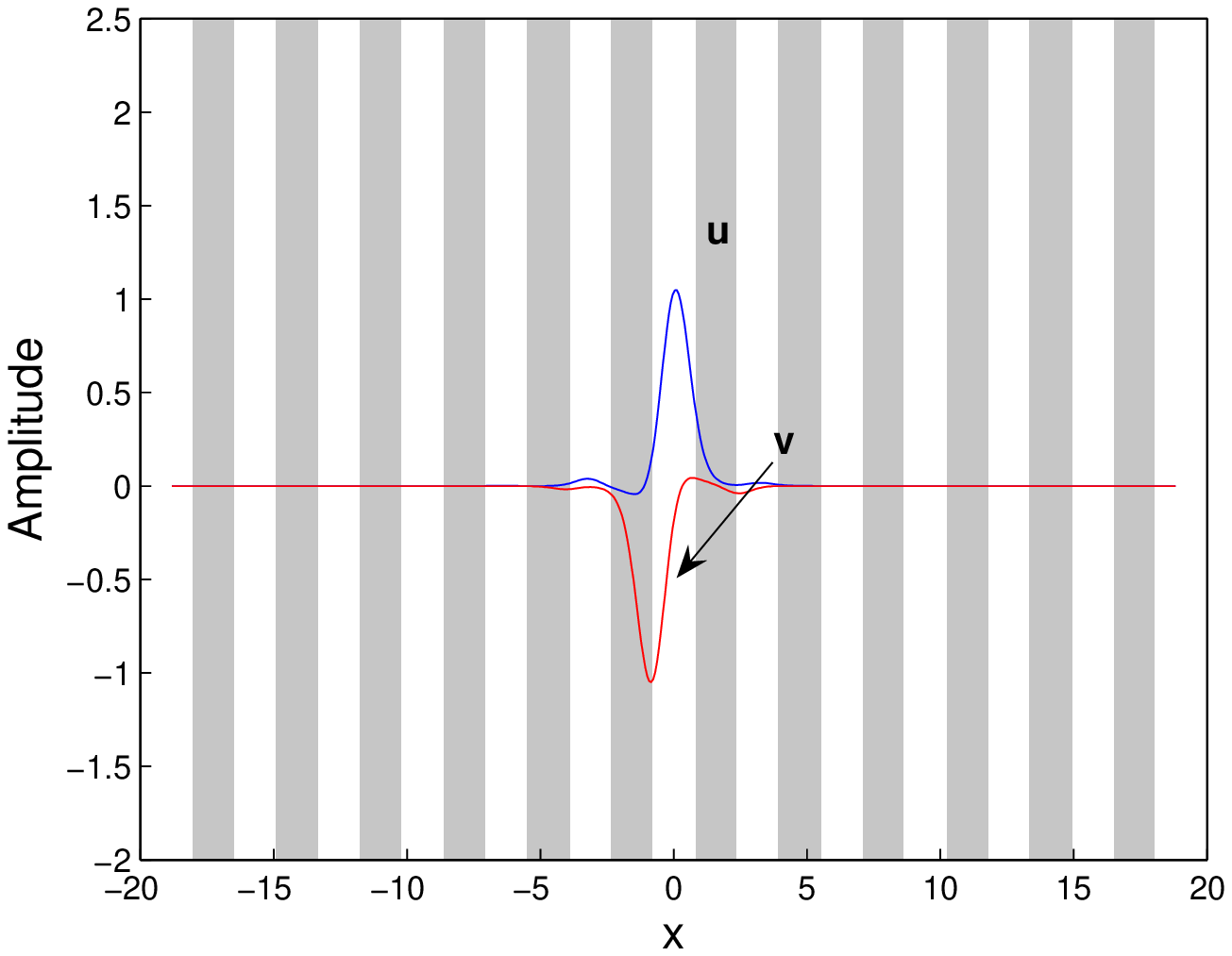}}
\subfigure[]{\includegraphics[width=3in]{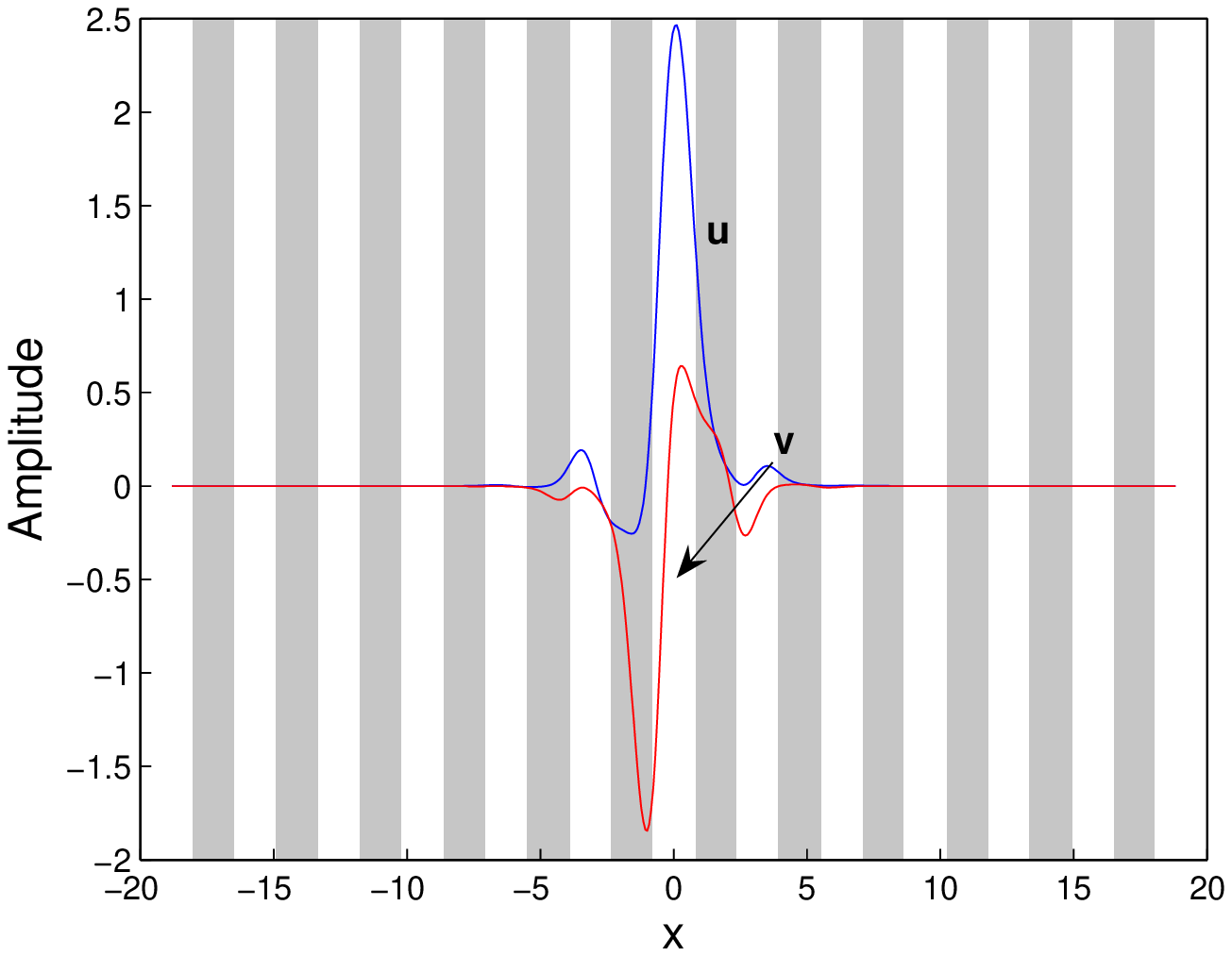}
\label{shifted00_2b}}
\subfigure[]{\includegraphics[width=3in]{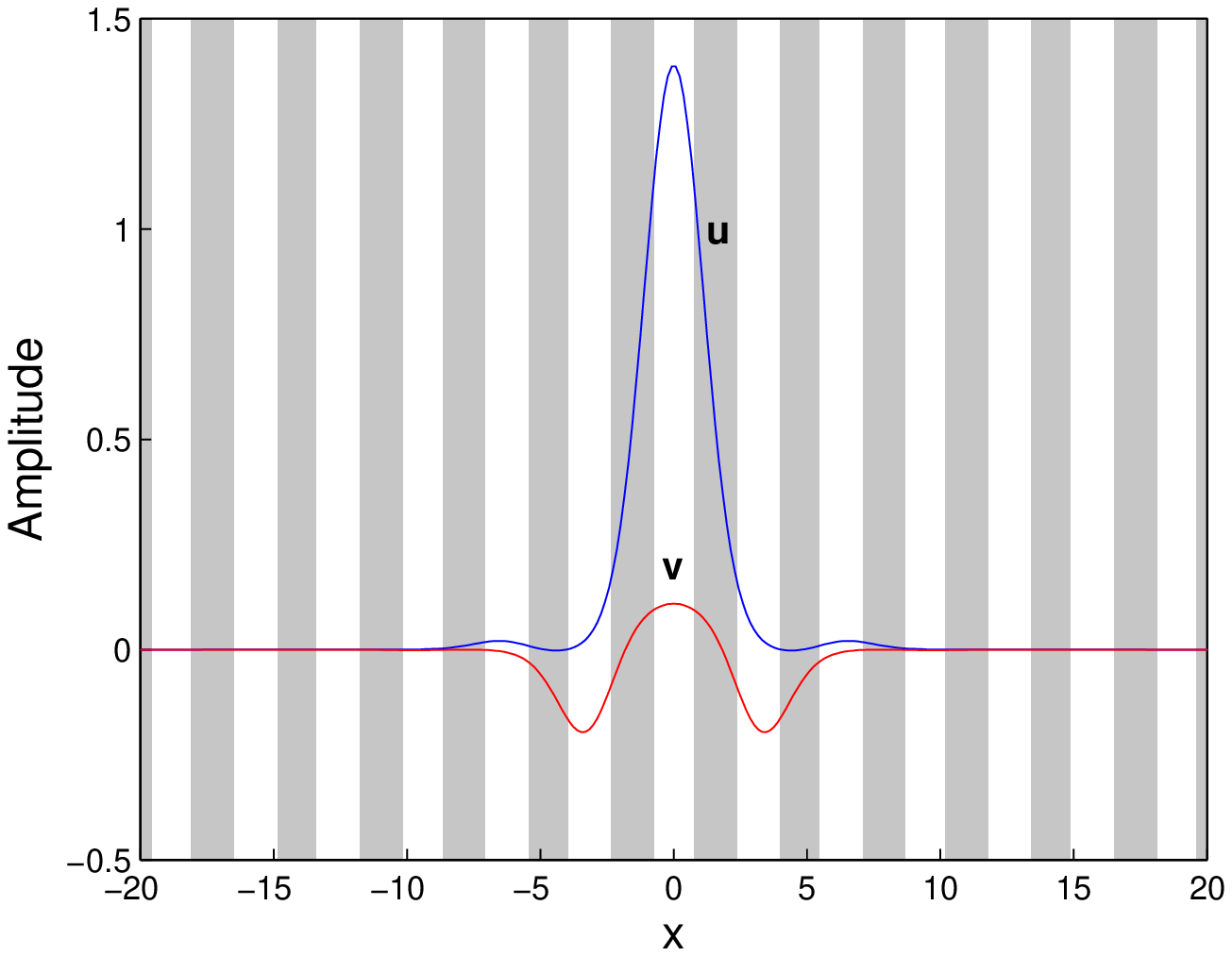}
\label{shifted00_2c}} \subfigure[]{%
\includegraphics[width=3in]{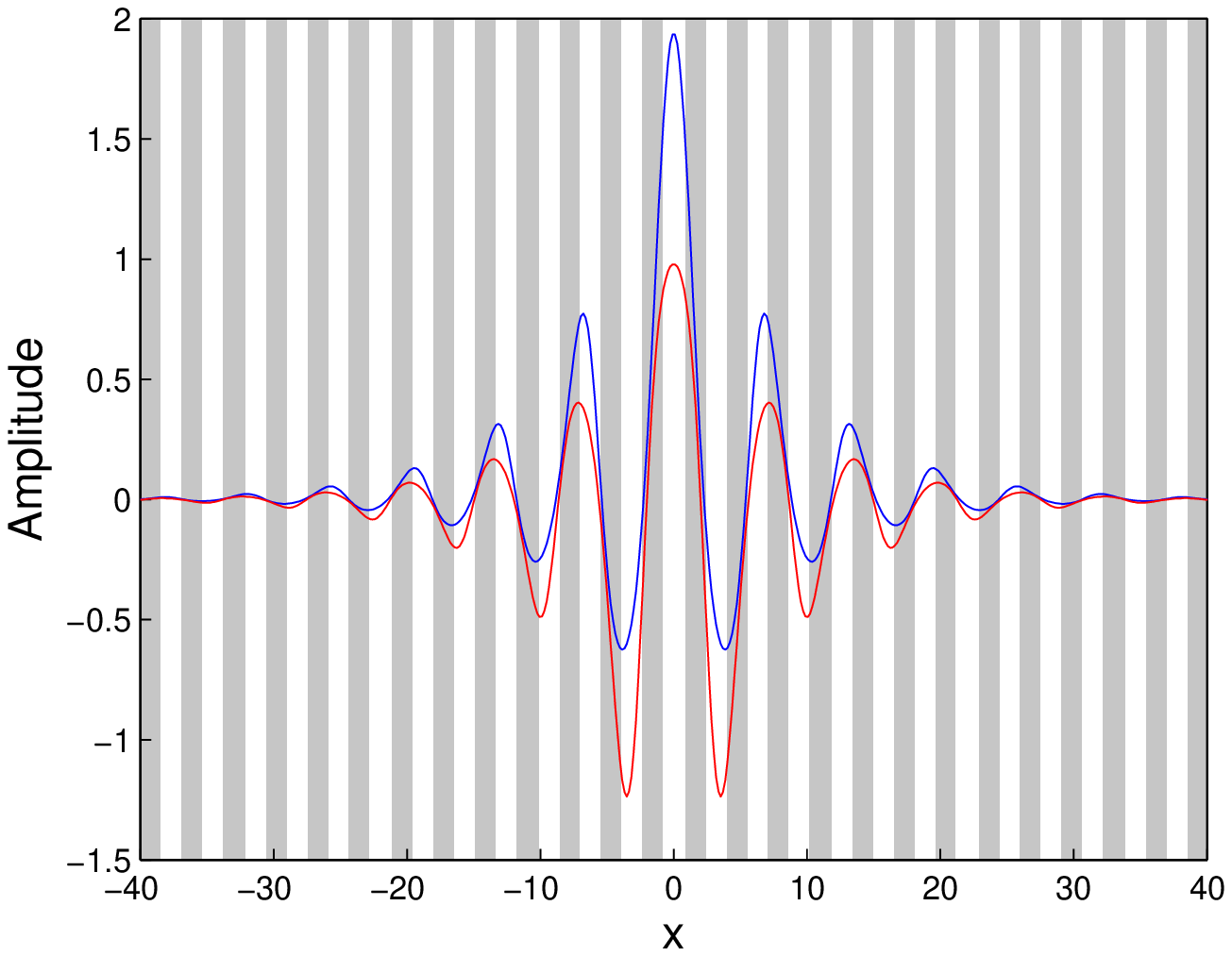}}
\caption{(Color online) Profiles of asymmetric solitons in the
mismatched repulsion-repulsion model with $\protect\varepsilon
=8$. (a) $\Delta =\pi /2$, $N=2$, $\Theta =0$, $\protect\mu
=-2.81$, $\protect\kappa =1.5$; (b) $\Delta =\protect\pi /2$,
$N=10$, $\Theta =0.31$, $\protect\mu =0.23$, $\protect\kappa
=2.5$; (c) $\Delta =\protect\pi $, $N=2$, $\Theta =0.92$,
$\protect\mu =-2.96$, $\protect\kappa =1$; (d) $\Delta =\pi$,
$N=10$, $\Theta =0.13$, $\protect\mu =-5.57$, $\protect\kappa
=8$.} \label{shifted00_2}
\end{figure}

Stability of the solitons in the mismatched AA and RR systems was studied in
direct simulations of Eqs. (\ref{model_1d}) over a wide range of values of $%
N $ and $\kappa $. In either case (as in the aligned system), the asymmetric
solitons were found to be stable whenever they exist, and the branch which
gives rise to the bifurcation is stable before the bifurcation and unstable
afterwards. Further, QAS and QS solitons in the AA and RR system,
respectively, are unstable at small $\varepsilon $ and stable at larger $%
\varepsilon $. When QS solitons are unstable in the AA system, they evolve
into their stable asymmetric counterparts. Similarly, in the misaligned RR
system, QAS solitons which were destabilized by the bifurcation (or which
are unstable due to the pseudo-bifurcation) transform themselves into stable
asymmetric solutions.

Soliton families in the mismatched system of the RA type were investigated
too. It was found that the corresponding picture is quite similar to that
presented for the RA system without mismatch in Fig. \ref{solitons1d_10}.

\section{Binary mixtures with linear coupling}

In this section we consider the model of a binary BEC in the single-core
trap equipped with an OL. As explained in Introduction, the two components
of the mixture ($\psi $ and $\phi $) represent two different spin states of
the same atom, and the linear coupling between them is induced by a
spin-flipping electromagnetic wave. The corresponding system of the
normalized GPEs for wave functions $\psi $ and $\phi $ takes the form%
\begin{equation}
\begin{array}{c}
i\psi _{t}+\psi _{xx}+\varepsilon \cos (2x)\psi +\left( g_{1}|\psi
|^{2}+g_{12}|\phi |^{2}\right) \psi +\kappa \phi =0, \\
i\phi _{t}+\phi _{xx}+\varepsilon \cos (2x)\phi +\left( g_{2}|\phi
|^{2}+g_{12}|\psi |^{2}\right) \phi +\kappa \psi =0%
\end{array}
\label{binary}
\end{equation}%
(in the single trap, there may be no mismatch between the lattice-potential
terms in the two equations). As the interaction between atoms may be both
attractive and repulsive, in this section we consider two basic cases: when
all the three nonlinear terms are attractive, i.e., $g_{1}=g_{2}\equiv 1$,$%
~g_{12}\geq 0$ (AAA), and when they all are repulsive, with $g_{1}=g_{2}=-1$%
, $g_{12}\leq 0$ (RRR).

Bifurcation diagrams for the AAA and RRR systems are shown in Fig. \ref%
{xpm11}. In the former case (AAA), stable asymmetric solitons bifurcate, as
before, from symmetric ones if $0\leq g_{12}<1$. However, if the
inter-species nonlinear interactions dominate, i.e., for $g_{12}>1$, the
character of the bifurcation changes, and asymmetric solitons bifurcate from
\emph{antisymmetric} ones. In the RRR system, a similar transition is
observed: the asymmetric solitons are generated, as above, by the
bifurcation from antisymmetric solitons if $0\leq -g_{12}<1$, but, for $%
-g_{12}>1$, antisymmetric solitons bifurcate from \emph{symmetric} ones. In
the \textit{Manakov's case}, $\left\vert g_{12}\right\vert =1$ \cite{Manakov}%
, no bifurcation was found. Note that linearly coupled equations (\ref%
{binary}) without the lattice potential ($\varepsilon =0$), but with the
linear coupling present, $\kappa \neq 0$, are equivalent to the Manakov's
integrable system with $\kappa =0$ \cite{Tratnik}, therefore the absence of
the bifurcation in this case is not surprising.
\begin{figure}[tbp]
\subfigure[]{\includegraphics[width=3in]{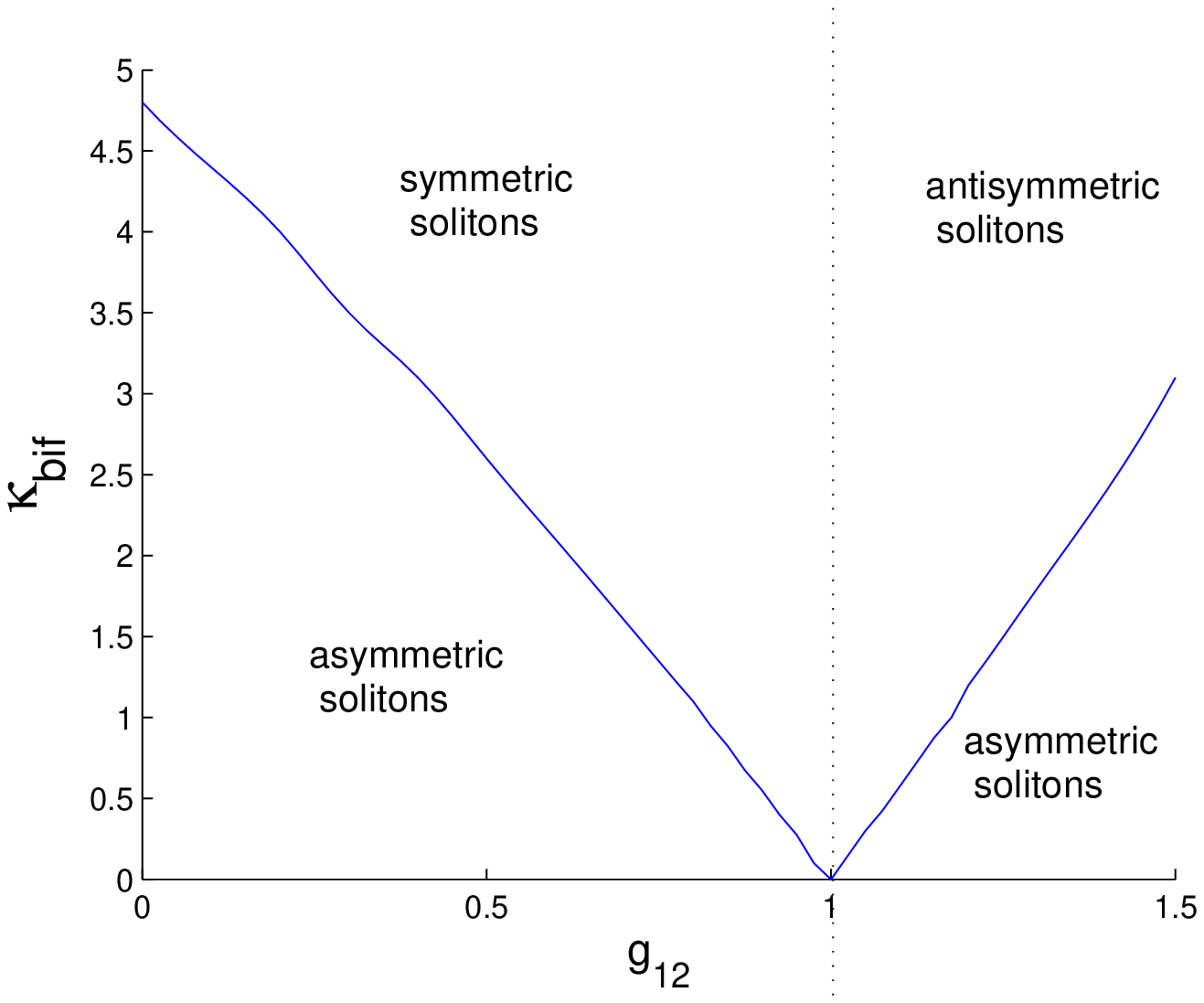}} \subfigure[]{%
\includegraphics[width=3in]{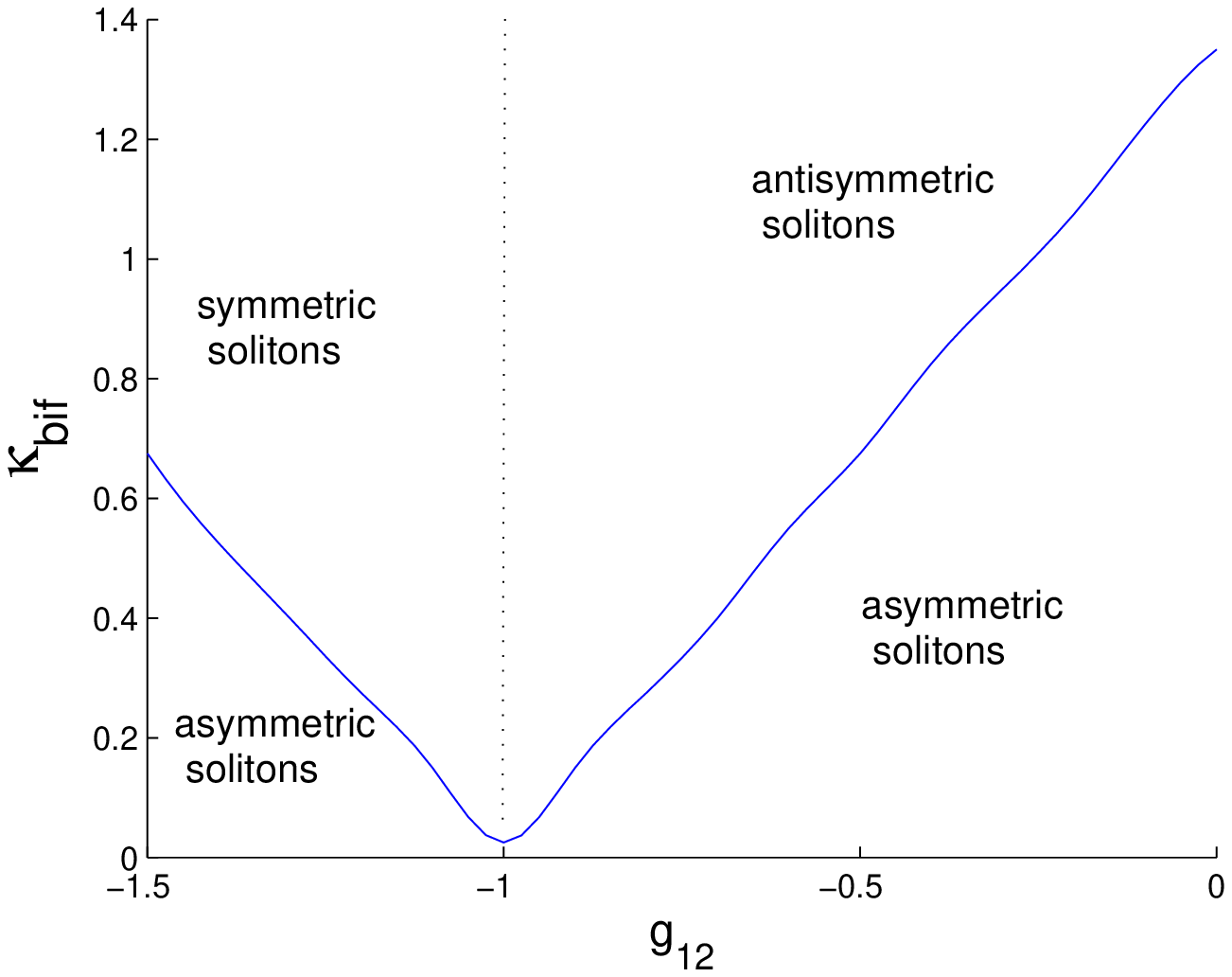}}
\caption{The bifurcation value of the coupling coefficient, $\protect\kappa %
_{\mathrm{bif}}$, versus the relative strength of the interspecies nonlinear
interaction, $g_{12}$, in the model of the binary mixture loaded into the
optical lattice with strength $\protect\varepsilon =8$ [Eqs. (\protect\ref%
{binary})]. The vertical dashed line divides regions in which asymmetric
solitons bifurcate from symmetric and antisymmetric ones. (a) The model with
attractive nonlinearities (for total norm $N=10$). In this system, the
asymmetric solitons are always stable. (b) The model with repulsive
nonlinearities (for $N=4$). In this case, all asymmetric solitons
bifurcating from the symmetric ones are stable, while the solitons
bifurcating from the antisymmetric branch are unstable if $\left\vert
g_{12}\right\vert $ is close to $1$ (but they are stable if $\left\vert
g_{12}\right\vert $ is small).}
\label{xpm11}
\end{figure}

Direct simulations of the AAA system demonstrate that the asymmetric
solitons, whenever they exist, are stable, and, as usual, the bifurcation
destabilizes the solitons from which the asymmetric ones emerge. In the RRR
system, the stability situation is more complex. While asymmetric solitons
bifurcating from the symmetric ones in the RRR system, i.e., in the case of $%
\left\vert g_{12}\right\vert >1$, are always stable, the asymmetric solitons
bifurcating, in the same system, from the antisymmetric branch (at $0\leq
\left\vert g_{12}\right\vert <1$) may be both stable and unstable. We did
not try to identify a border between the stable and unstable asymmetric
solitons in the latter case in an accurate form; however, the asymmetric
solitons in the RRR system are definitely stable for $\left\vert
g_{12}\right\vert \rightarrow 0$ (which complies with the results reported
above for the RR system with $g_{12}=0$), and they are unstable for $%
\left\vert g_{12}\right\vert $ close to (but smaller than) $1$. In that
case, direct simulations demonstrate that unstable asymmetric solitons
transform into persistent breathers (not shown here). We did not
systematically study the stability of the nonbifurcating families, i.e.,
antisymmetric and symmetric ones in the AAA and RRR systems, respectively.

\section{Conclusion}

In this work, we have studied families of soliton states in a symmetric set
of two effectively one-dimensional traps (cores) equipped with OLs (optical
lattices), filled with self-attractive or self-repulsive BEC, and coupled by
linear tunneling (the same model may be also realized in terms of spatial
solitons in two parallel planar optical waveguides, which carry lattices in
the form of a transverse modulation of the refractive index). Asymmetric
systems, with a phase shift $\Delta $ between the two OLs, as well as with
opposite signs of the nonlinearity in the BEC confined in the different
traps, were considered too. In the spectrum of the dual-core system, the
linear coupling between the cores splits finite bandgaps induced by the
single OL into subgaps, or partly closes them. Influence of mismatch $\Delta
$ on the spectrum was considered too, with a conclusion that large $\Delta $
tends to make the spectrum qualitatively similar to that in the single-core
model. In the full nonlinear model, families of symmetric and antisymmetric
solitons, as well as asymmetric solitons generated by symmetry-breaking
bifurcations, have been constructed. This was done in the case of attraction
or repulsion in both cores (AA and RR systems), as well as in the mixed (RA)
system. Deformation of the soliton families in the AA and RR system under
the action of mismatch $\Delta $ between the two OLs was investigated too.

In the AA and RR systems, asymmetric solitons bifurcate supercritically from
symmetric and antisymmetric ones, respectively. In the former case,
symmetric and asymmetric fundamental solitons were found only in the
semi-infinite gap, while antisymmetric solitons were also discovered in the
first finite subgap, as well in the Bloch band separating it from the
semi-infinite band (in the latter case, the antisymmetric solitons have the
character of \textit{embedded} ones). In the RR system, solitons are located
in finite bandgaps, and, additionally, antisymmetric and symmetric solitons
were found as embedded ones inside Bloch bands. In both systems (AA and RR),
the asymmetric solitons are always stable, while the branch which gives rise
to them is stable before the bifurcation, and unstable afterwards. The
nonbifurcating branches, i.e., antisymmetric and symmetric solutions in the
AA and RR systems, respectively, are unstable or stable as a whole in weak
and strong OL, respectively. The corresponding stability region was
identified in a complete form for the AA system, in the parameter plane of
coupling coefficient $\kappa $ and OL strength $\varepsilon $. In the
systems of both AA and RR types, bistability regions were found, in which
the stable asymmetric solitons coexist with either antisymmetric or
symmetric ones, respectively. If the symmetric soliton in the AA system, or
the antisymmetric soliton in the RR system was destabilized by the
bifurcation, its evolution tends to transform it into a stable asymmetric
soliton, in either system. In the RA system, two soliton families were
found, dominated by either the self-attractive or repulsive component, both
being entirely stable.

In the AA system, we have also studied families of twisted (odd) solitons,
featuring two out-of-phase peaks in a cell of the OL. The behavior of these
families is, generally, similar to that of the fundamental solitons,
including the bifurcation of the symmetric branch into an asymmetric one,
and bistability; however, unlike the situation with the fundamental
solitons, the symmetry-breaking bifurcation of the twisted solitons occurs
in a finite bandgap. In addition, $\pi $-out-of-phase bound states of
fundamental solitons were constructed in both the AA and RR systems. Their
behavior too is quite similar to that of the respective fundamental
solitons, with a difference that, past the bifurcation, an unstable bound
state tends to arrange itself into a persistent breather (rather than into a
stationary bound state of the asymmetric type).

The study of the systems with the mismatch, $\Delta $, between the OLs in
the two cores demonstrates that former symmetric and antisymmetric solitons
turn into QS (quasi-symmetric) and QAS (quasi-antisymmetric) ones. For $%
\Delta =\pi /2$, the deformation is mild, and the behavior of various
soliton families remains qualitatively the same as in the system with $%
\Delta =0$; in particular, well-defined bifurcations which generate
asymmetric solitons from QS or QAS ones, in the mismatched AA and RR system,
respectively, occur despite the presence of the mismatch (in a system of
mismatched parallel-coupled Bragg gratings, similar observations were
recently made in Ref. \cite{Yossi}). However, when the mismatch attains its
maximum, $\Delta =\pi $, the situation becomes different: instead of the
bifurcations, both the AA and RR systems demonstrate \textit{%
pseudo-bifurcations}, when the branch of asymmetric solutions does not
really bifurcate from that of QS or QAS solutions, but rather goes very
close and only asymptotically merges into it. Thus, the asymmetric branch
always exists and is stable in the latter case, while its QS or QAS
counterpart is always unstable (although the instability is extremely weak
when it approaches the stable asymmetric branch). To the best of our
knowledge, the present work reports the first example of the
pseudo-bifurcation in a mismatched binary system (in a system of mismatched
Bragg gratings, no such effect was observed \cite{Yossi}).

Finally, we have considered a model of the single-core trap, equipped with
the OL and filled with a binary mixture, assuming both the nonlinear
interaction between the two species, and linear coupling between them,
induced by a resonant spin-flipping field, if the species correspond to
different spin states of the same atom. In this case, we have found that the
bifurcations generating asymmetric solitons from symmetric and antisymmetric
ones are qualitatively similar to their counterparts in the model without
the nonlinear interaction between the species, if this interaction is weaker
than the intra-species nonlinearity; otherwise, the bifurcations switch
their character, so that the asymmetric solitons emerge from antisymmetric
and symmetric ones in the AA and RR systems, respectively.

A very natural extension would be to consider a two-dimensional version of
the models introduced in this work, which may also directly apply to BEC in
a dual-core pancake-shaped trap, or to a binary mixture in a single-core
nearly flat trap. The work in this direction is currently in progress, and
results will be reported elsewhere. It may also be possible to consider the
spontaneous symmetry breaking in a degenerate fermion gas filling a
dual-core trap, using a relatively simple description of the fermion gas
based on the Thomas-Fermi (alias mean-field-hydrodynamic) approximation,
which was developed, for various settings, in Refs. \cite{hydro}.

\section*{Acknowledgement}

This work was supported, in a part by the Israel Science Foundation through
the Center-of-Excellence grant No. 8006/03.

\end{document}